\documentclass[sigconf]{acmart}
\usepackage{graphicx}
\usepackage{array}
\usepackage{amsmath}
\usepackage{algorithm}
\usepackage{multirow}
\usepackage{subfigure}
\usepackage{caption}
\usepackage{colortbl}
 \usepackage{enumitem}
 \usepackage{balance}
\usepackage{fancyhdr}

\usepackage{float}
\usepackage{graphicx}
\usepackage{listings}
\usepackage{color}
\definecolor{codegreen}{rgb}{0,0.6,0}
\definecolor{codegray}{rgb}{0.5,0.5,0.5}
\definecolor{codepurple}{rgb}{0.58,0,0.82}
\definecolor{backcolour}{rgb}{0.95,0.95,0.92}
\definecolor{dkgreen}{rgb}{0,0.6,0}
\definecolor{gray}{rgb}{0.5,0.5,0.5}
\definecolor{mauve}{rgb}{0.58,0,0.82}
\definecolor{tablegreen}{rgb}{0.32,0.65,0.36}
\definecolor{codegray}{rgb}{0.9,0.9,0.9}
\lstdefinestyle{mystyle}{
 language=Java,
 showstringspaces=false,
 columns=flexible,
 basicstyle={\ttfamily},
 numbers=none,
 numberstyle=\color{gray},
 keywordstyle=\color{blue},
 commentstyle=\color{dkgreen},
 stringstyle=\color{dkgreen},
 breakatwhitespace=true,
}

\lstdefinestyle{mystyle_python}{
 language=Python,
 showstringspaces=false,
 columns=flexible,
 basicstyle={\ttfamily},
 numbers=none,
 numberstyle=\color{gray},
 keywordstyle=\color{blue},
 commentstyle=\color{dkgreen},
 stringstyle=\color{dkgreen},
 breakatwhitespace=true,
}

\newcommand{\tool}{DECOM}

\makeatletter

\copyrightyear{2022} 
\acmYear{2022} 
\setcopyright{rightsretained} 
\acmConference[ASE '22]{37th IEEE/ACM International Conference on Automated Software Engineering}{October 10--14, 2022}{Rochester, MI, USA}
\acmBooktitle{37th IEEE/ACM International Conference on Automated Software Engineering (ASE '22), October 10--14, 2022, Rochester, MI, USA}
\acmDOI{10.1145/3551349.3556917}
\acmISBN{978-1-4503-9475-8/22/10}

\begin{document}

% \title{Improving Comment Generation with Multi-Pass Rewriting}
% \title{Rewrite and Evaluate: A Novel Iterative Method for Comment Generation}
% \title{Comment Generation via Iterative Rewriting and Evaluation}
% \title{Iterative Multi-Task Learning for Neural Comment Generation}
\title{Automatic Comment Generation via Multi-Pass Deliberation}
\author{Fangwen Mu}
\authornote{Also With Laboratory for Internet Software Technologies, Institute of Software, CAS}
\authornote{Also With University of Chinese Academy of Sciences}
\email{fangwen2020@iscas.ac.cn}
\affiliation{%
\institution{Institute of Software, Chinese Academy of Sciences}
\city{Beijing}
\country{China}
}

\author{Xiao Chen}
\authornotemark[1]
\authornotemark[2]
\email{chenxiao2021@iscas.ac.cn}
\affiliation{%
\institution{Institute of Software, Chinese Academy of Sciences}
\city{Beijing}
\country{China}
}

\author{Lin Shi}
\authornotemark[1]
\authornotemark[2]
\authornote{Corresponding author}
\email{shilin@iscas.ac.cn}
\affiliation{
\institution{Institute of Software, Chinese Academy of Sciences}
\city{Beijing}
\country{China}
}

\author{Song Wang}
\email{wangsong@eecs.yorku.ca}
\affiliation{%
  \institution{Lassonde School of Engineering, York University}
  \city{Toronto}
  \country{Canada}
}

\author{Qing Wang}
\authornotemark[1]
\authornotemark[2]
\authornotemark[3]
\authornote{Also With Science \& Technology on Integrated Information System Laboratory, Institute of Software, CAS}
\email{wq@iscas.ac.cn}
\affiliation{
\institution{Institute of Software, Chinese Academy of Sciences}
\city{Beijing}
\country{China}
}

% \author{Fangwen Mu$^{1,3}$, Xiao Chen$^{1,3}$,  Lin Shi$^{1,3}$,Song Wang$^{4}$, Qing Wang$^{1,2,3}$}
% \author{Fangwen Mu$^{1,3}$, Xiao Chen$^{1,3}$,  Lin Shi$^{1,3}$, Song Wang$^{4}$, Qing Wang$^{1,2,3}$}
% \affiliation{$^1$Laboratory for Internet Software Technologies, Institute of Software Chinese Academy of Sciences
% \city{Beijing}
% \country{China}}
% \affiliation{$^2$ State Key Laboratory of Computer Science, Institute of Software Chinese Academy of Sciences
% \city{Beijing}
% \country{China}}
% % \affiliation{ $^1$Laboratory for Internet Software Technologies,}
% % \affiliation{$^2$ State Key Laboratory of Computer Science, \\Institute of Software Chinese Academy of Sciences
% % \city{Beijing}
% % \country{China}}
% \affiliation{$^3$ University of Chinese Academy of Sciences
% \city{Beijing}
% \country{China}}
% \affiliation{$^4$ York University, Lassonde School of Engineering
% \country{Canada} \\
% \{fangwen2020,chenxiao2021,shilin,wq\}@iscas.ac.cn,
% wangsong@eecs.yorku.ca
% }
% % \affiliation{$^*$ Corresponding author.}
% \authornote{Corresponding author.}

%author：Fangwen Mu, Xiao Chen, Lin Shi, Song Wang, Qing Wang
\begin{abstract}

Deliberation is a common and natural behavior in human daily life. For example, when writing papers or articles, we usually first write drafts, and then iteratively polish them until satisfied.
In light of such a human cognitive process, we propose {\tool}, which is a multi-pass deliberation framework for automatic comment generation. 
{{\tool} consists of multiple \emph{Deliberation Models} and one \emph{Evaluation Model}.
Given a code snippet, we first extract keywords from the code and retrieve a similar code fragment from a pre-defined corpus. Then, we treat the comment of the retrieved code as the initial draft and input it with the code and keywords into {\tool} to start the iterative deliberation process. 
At each deliberation, the deliberation model polishes the draft and generates a new comment. The evaluation model measures the quality of the newly generated comment to determine whether to end the iterative process or not. When the iterative process is terminated, the best-generated comment will be selected as the target comment.}
Our approach is evaluated on two real-world datasets in Java (87K) and Python (108K), and experiment results show that our approach 
outperforms the state-of-the-art baselines. 
A human evaluation study also confirms the comments generated by {\tool} tend to be more readable, informative, and useful. 

\end{abstract} 

\begin{CCSXML}
<ccs2012>
   <concept>
       <concept_id>10011007.10011006</concept_id>
       <concept_desc>Software and its engineering~Software notations and tools</concept_desc>
       <concept_significance>500</concept_significance>
       </concept>
    <concept>
       <concept_id>10010147.10010178</concept_id>
       <concept_desc>Computing methodologies~Artificial intelligence</concept_desc>
       <concept_significance>500</concept_significance>
       </concept>
 </ccs2012>
\end{CCSXML}

\ccsdesc[500]{Software and its engineering~Software notations and tools}
\ccsdesc[500]{Computing methodologies~Artificial intelligence}

\keywords{Comment generation, Information retrieval, Deep neural network}

\maketitle

\section{Introduction}
\label{sec:introduction}

With software growing in size and complexity, developers tend to averagely spend around 59\% of their effort on program comprehension during software development and maintenance \cite{DBLP:conf/icse/XiaBLXHL18, DBLP:journals/tse/KoMCA06}. 
Source code comments provide concise natural language descriptions of code snippets, which not only greatly reduce the effort for developers to understand the code, but also play a vital role in software maintenance and evolution. 
However, manually commenting code is time-consuming, and code comments are often missing or outdated in software projects \cite{DBLP:conf/sigdoc/SouzaAO05, DBLP:journals/ese/Kajko-Mattsson05}. 
Therefore, the code comment generation task, which aims at automatically generating a high-quality comment for a given code snippet, has long attracted the interest of many researchers. 

% Code comment generation concerns the production of a concise and fluent description of source code that facilitates software development and maintenance by enabling developers to comprehend, ideate, and document code effectively.
Most of the existing approaches treat comment generation as a machine translation task and adopt a one-pass encoder-decoder process, i.e., first encode the input code into a sequence of semantic features, then decode the features to a natural language comment~\cite{DBLP:conf/acl/IyerKCZ16,DBLP:conf/iwpc/HuLXLJ18,DBLP:conf/icse/ZhangW00020,DBLP:conf/kbse/LiL000J21}.
Although the encoder-decoder framework has achieved {remarkable performance} on the comment generation task, it still suffers from two major limitations. 
The first one is that such models adopt a regular one-pass decoding process {that sequentially generates comments word by word}. They directly use the generated comment as the final output, which results in their inability to correct the mispredicted words. The words mistakenly predicted in the early steps may lead to error accumulation under the constraint of the language model.
%\song{may need to mention early about  they generate comments word\-by\-word sequentially}
\begin{figure}[t]
% \vspace{-0.4cm}
\centering
\includegraphics[width=\columnwidth]{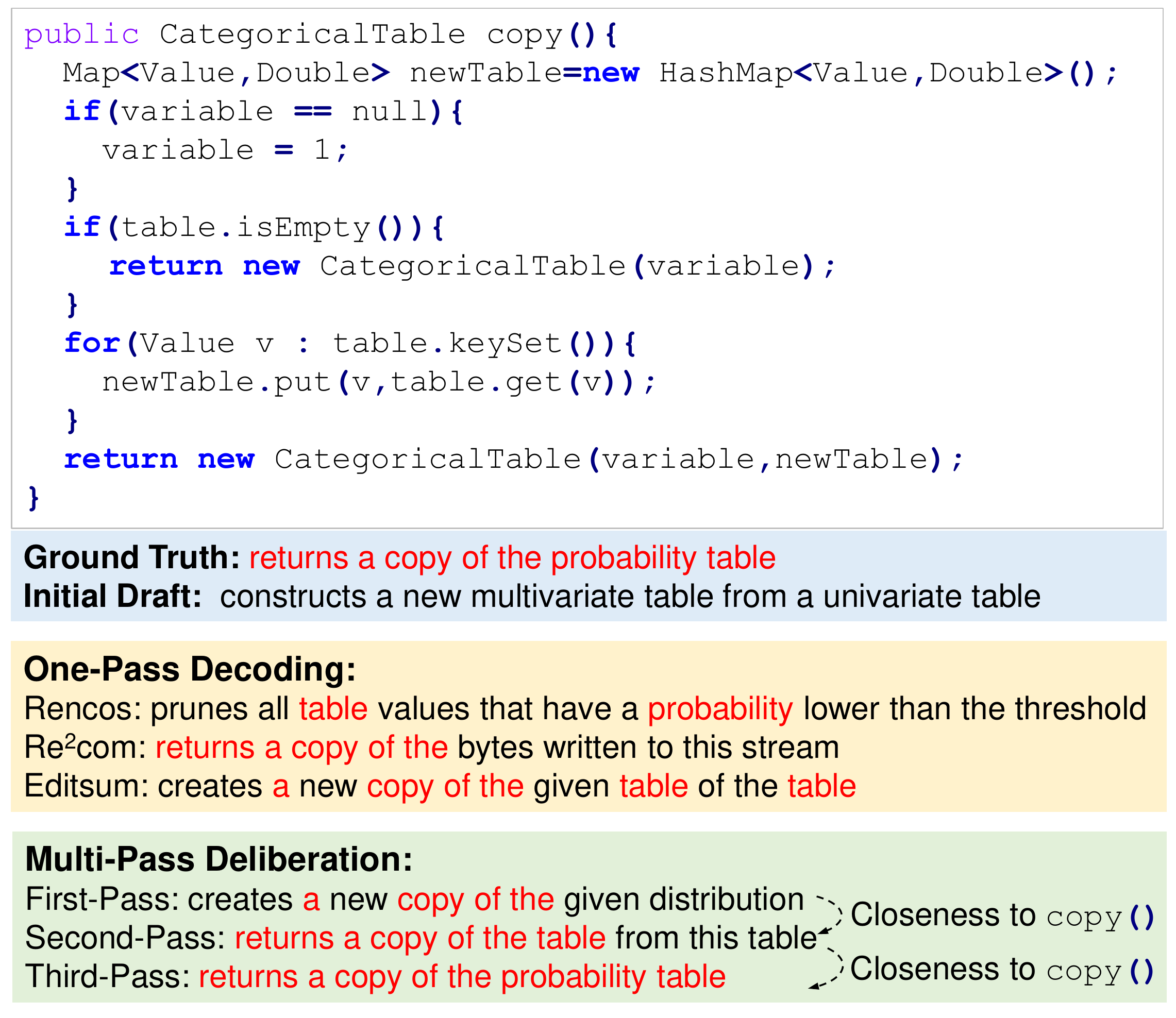}
\vspace{-0.6cm}
\caption{A motivation example of multi-pass deliberation.}
%\song{replace with a high resolution image, current one is not clear when print out}}
\vspace{-0.5cm}
\label{fig:motivation}
\end{figure}
Taking the auto-generated comments in Figure \ref{fig:motivation} as an example, the one-pass model Re$^2$com \cite{DBLP:conf/kbse/WeiLLXJ20} incorrectly predicts the sixth word ``probability'' as ``bytes'', which leads the model to keep exploring the words related to ``bytes'' when predicting the consequential words. As a result, the related words ``written to this stream'' are mistakenly generated, which
results in a typical example of error accumulation. 
The second limitation is that they generate comments sequentially.
% meaning that when the decoder predicts a certain word, the global information is not considered
%The second limitation is that they generate comments in a left-to-right manner\song{can we say "sequentially"?}, 
%meaning that when the decoder predicts the $i$-th word $y_i$, only left words $y_{<i}$ are considered, while the right words $y_{>i}$ cannot be utilized. 
% \song{readers will be confused on this, as comments were not available when generating, might need to revise or directly say the global information is not considered xxx} 
Thus, such {sequential} one-pass models cannot leverage the global information of the generated comment to further polish its local content. As the example shown in Figure \ref{fig:motivation}, the one-pass model EditSum \cite{DBLP:conf/kbse/LiL000J21} generates two consecutive prepositional phrases after the word ``cop''. Although either of them is reasonable in their local contexts (``of the given table'' and ``of the table''), putting them together degrades the comment fluency and makes the developers hard to understand.

To alleviate these challenges, we introduce the \emph{Deliberation} mechanism~\cite{DBLP:conf/nips/XiaTWLQYL17} in the comment generation task, aiming to further enhance the performance. 
Deliberation is a common and natural behavior in human daily life. When writing papers or articles, we usually first write drafts, and then iteratively polish them until satisfied. 
{Figure \ref{fig:motivation} illustrates an example of applying multi-pass deliberation on comment generation.
Based on the initial draft ``constructs a new multivariate table from a univariate table'', the first-pass deliberation will generate the comment ``creates a new copy of the given distribution'', and refine it in the second and the third pass guided by the closeness to the similarity with the give code snippet. 
In the end, we could obtain the most satisfying comment ``returns a copy of the probability table''.}

%As shown in Figure \ref{fig:motivation}, the initial draft and the keywords are extracted by using the given code. For the first-pass generation, {\tool} refines the initial draft to the comment ``calculates the value from a given log'', and calculates the quality score of it. For the second pass, {\tool} realizes the two words ``value'' and ``log'' are in the wrong place by utilizing the global information of the comment, and continues polishing the comment to generate a new comment, which is exactly the same as the ground truth. The quality score of this comment is larger than the first-pass generated comment, so we reserve it. For the third-pass, {\tool} mistakenly predicts a word, and found the quality score of it is lower than the previous one. Ultimately, {\tool} selects the second-pass generated comment as the target comment. \song{this part can be more concise, too much details}

In light of such a human cognitive process, we propose a novel multi-pass deliberation framework for automatic comment generation, named {\tool}, which contains multiple \emph{Deliberation Models} and one \emph{Evaluation Model}. Given a code snippet, initially, we retrieve the most similar code from a pre-defined corpus and treat its comment as the initial draft. {We also extract the identifier names from the input code as keywords, since these user-defined words usually contain more semantic information that users want to express \cite{DBLP:journals/jpl/TakangGM96,DBLP:conf/csmr/ButlerWYS10,DBLP:conf/iwpc/SchankinBHHRB18}. Then, we input the code, the keywords, and the initial draft into {\tool} to start the iterative deliberation process. At each deliberation, the deliberation model polishes the draft and generates a new comment. The evaluation model calculates the quality score of the newly generated comment. This multi-pass process terminates when (1) the quality score of the new comments is no longer higher than the previous ones, or (2) the maximum number of deliberations is reached.}
To evaluate our approach, we conduct experiments on two real-world datasets in Java (87K) and Python (108K), and the results show that our approach outperforms the state-of-the-art baselines by 8.3\%, 6.0\%, 13.3\%, and 10.5\% with respect to BLEU-4, ROUGE-L, METEOR, and CIDEr on Java dataset. On Python dataset, {\tool} improves the performance on BLEU-4, ROUGE-L, METEOR, and CIDEr by 5.8\%, 3.8\%, 6.6\%, and 6.3\%, respectively. We also conduct a human evaluation to assess the generated comments on three aspects: naturalness, informativeness, and usefulness, showing that {\tool} can generate useful and relevant comments. 

Our main contributions are outlined as follows:
\begin{itemize}%[leftmargin=*]
    \item \textbf{Technique}: a multi-pass deliberation framework for comment generation, named {\tool}, which is inspired by the human cognitive process, and can effectively generate comments in an iterative way. To the best of our knowledge, this is the first work that employs multi-pass deliberation to enhance the performance of comment generation.
    \item \textbf{Evaluation}: an experimental evaluation of the performance of {\tool} against state-of-the-art baselines, which shows that {\tool} outperforms all baselines, together with a human evaluation, which further confirms the readability, informativeness, and usefulness of {\tool}.
    \item \textbf{Data}: publicly accessible dataset and source code~\cite{website} to facilitate the replication of our study and its application in extensive contexts.
\end{itemize} 

In the rest of this paper, 
Section 2 elaborates the approach. 
Section 3 presents the experimental setup. 
Section 4 demonstrates the results and analysis. 
Section 5 describes the human evaluation.
Section 6 discusses indications and threats to validity. 
Section 7 introduces the related work. 
Section 8 concludes our work.

%which contains multiple deliberation models and one evaluation model. The deliberation models improve the previously generated comments by utilizing their global information, and the evaluation model estimates the quality of the new generated comments.
% \input{sec/2.background}
\section{Approach}
\label{sec:method}

\begin{figure*}[tbh]
\centering
\includegraphics[width=\textwidth]{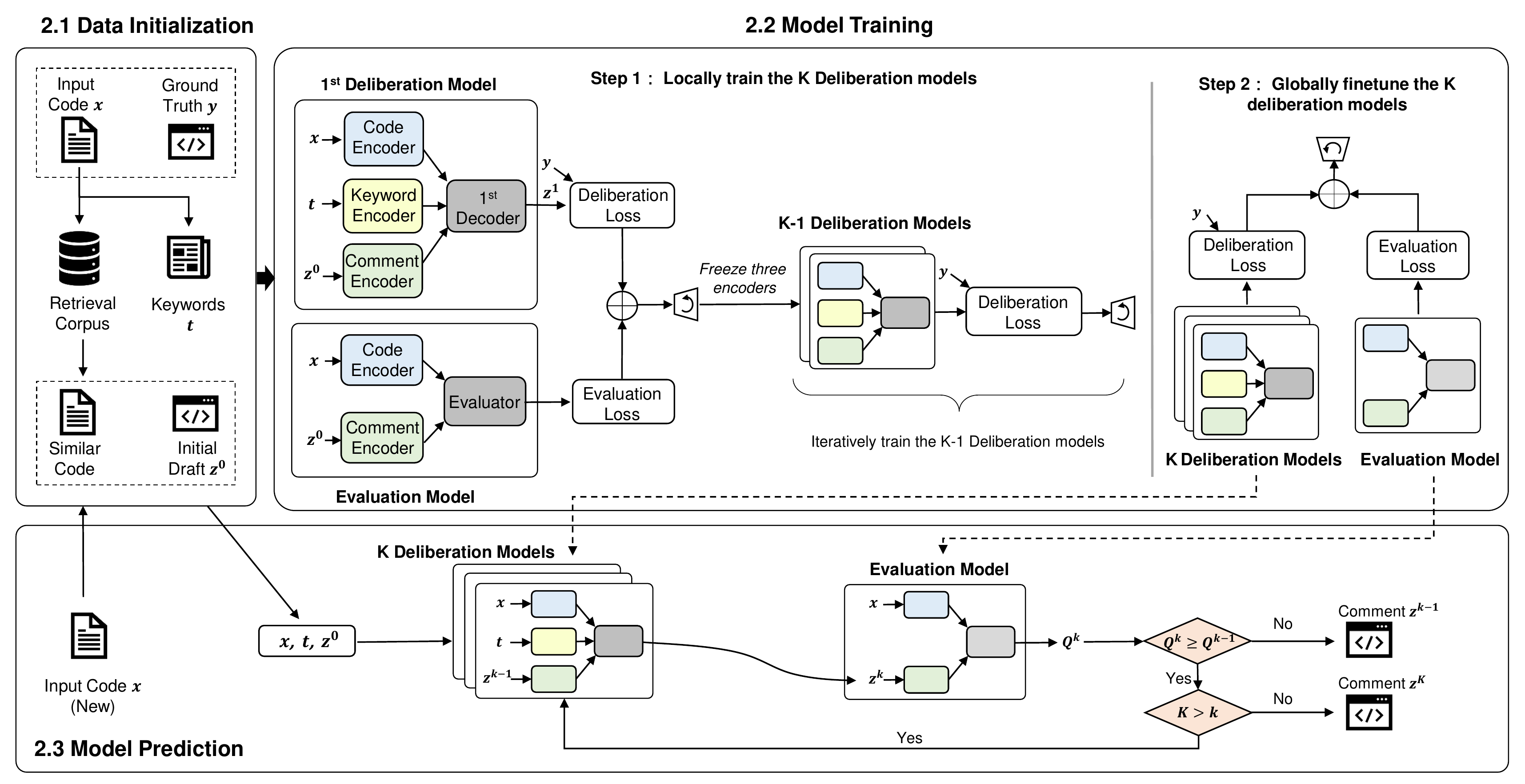}
\caption{The overall architecture of {\tool}}
% \mfw{need update, shared encoder, symbols}%\song{seems this only contains the training process, should we also show how it works for a given query?}.}
\label{fig:framework}
\vspace{-0.2cm}
\end{figure*}
% The comment generation process involves multiple passes: for each turn, the Rewrite module generates a new candidate comment given the source code, the keyword, and the candidate comment generated by the prior pass. The evaluate module calculates the similarity score between new candidate comment and source code, and stores the score. Finally, the comment with the highest score will be selected as the target comment.
In this section, we present our {\tool}, a multi-pass deliberation framework that performs an iterative polishing process to refine the draft to a better comment.
Figure \ref{fig:framework} illustrates an overview of {\tool}, which consists of three main stages:
(1) \textbf{Data initialization}, for extracting the keywords from the input code and retrieving the similar code-comment pair from the retrieval corpus;
(2) \textbf{Model training}, for leveraging a two-step training strategy to optimize {\tool};
% that contains multiple deliberation models and one evaluation model;
% constructing our automatic comment generation framework that contains $K$ Deliberation models and an Evaluation model.
% (3) \textbf{model training}, for optimizing the parameters of {\tool} by leveraging a two-step training strategy.
and (3) \textbf{Model prediction}, for generating the target comment of the new source code. 
Below, we provide details for each stage in {\tool}.

% At each turn, {\tool} first input the past draft into the Rewrite module to polish it and generate an updated comment. Then, {\tool} input the source code and the updated comment into the Evaluate module to calculate their similarity as the quality score of the updated comment. Termination of this multi-pass deliberation process depends on whether the quality score of the updated comments are no longer higher than the previous ones.

\subsection{Data Initialization}
Given a code snippet $x$, this stage aims to extract the keywords $t$ {from $x$}, and retrieve the initial draft $z^0$ from the retrieval corpus.

\textbf{Extract keywords from code}. {A code snippet contains many different types of tokens, such as reserve words (\emph{if}, \emph{for}), identifier names (\emph{set\_value}, \emph{SortList}), and operators ($+$, $*$). Among them, identifier names defined by users usually contain more semantic information that users want to express \cite{DBLP:journals/jpl/TakangGM96,DBLP:conf/csmr/ButlerWYS10,DBLP:conf/iwpc/SchankinBHHRB18}. For example, a method’s name is a typical identifier name, which is used to describe the overall functionality of the code and can be considered as a shorter version of its code comment. Thus, to enable the model to attend more on the identifier names and capture semantic information from them, we extract these words from code.
First, we use the \textit{javalang} \cite{javalang} and \textit{tokenize} \cite{tokenize} libraries to extract identifier names from the Java and Python code snippets, respectively. Then, we further split the extracted identifier names into sub-tokens by \emph{CamelCase} or \emph{snake\_case} to obtain the smaller semantic units and reduce data sparsity. These sub-tokens are treated as the keywords of the code.}

\textbf{Retrieve the initial draft.}
{To obtain the initial draft, following the previous studies \cite{DBLP:conf/kbse/WeiLLXJ20}, we use the lexical similarity-based retrieval method to identify the top similar code-comment pair for the given code $x$.} 
%\lin{kind of confuse why give a sentence like this here, it does not give an overall summarization for the following points. suggest to move behind, and just start with first, second,...}
Specifically, we first take the training set of the benchmark dataset as the retrieval corpus. Then, for each code in the retrieval corpus, we adopt the BM25 \cite{DBLP:journals/ftir/RobertsonZ09} metric to calculate the similarity between it and the given code $x$. The BM25 is a bag-of-words retrieval metric to measure the relevance of documents to a given search query in IR and is also widely used in code clone detection and code search tasks \cite{DBLP:journals/tsc/JiangNSRKZL19,DBLP:conf/pldi/SachdevLLKS018,DBLP:journals/corr/abs-2203-07722}. Finally, we extract the code with the highest similarity score as the retrieved result,
% . (note that to avoid exactly duplication, we select the second-ranked code when training)\lin{need rewrite}, 
and use the comment of the code as the initial draft $z^0$.
%\song{do we have a similarity threshold about selecting z? if the largest similarity is extremely low, say 0.1 are we still use it?} \mfw{Yes, we will still use it. we do not have a threshold, since we must have a initial draft to start polishing no matter its quality. But in the next submission, we can add an experiment to explore the impact of different qualities of the initial draft on model performance. Or use some strategy to improve this part}. 
Since the size of our training sets is quite large, we leverage the open-source search engine Lucene \cite{lucene} to speed up the retrieval process. We follow the settings of Lucene from Re$^2$Com \cite{DBLP:conf/kbse/WeiLLXJ20} to run our experiments. 
%\lin{The last sentence seems straggling with the retrieval process.}

\subsection{Model Training}
{\tool} contains $K$ deliberation models and one evaluation model, where $K$ is the maximum deliberation number.
% , where the $k^{th}$ Deliberation model performs the $k^{th}$ polishing process to improve the ${(k-1)}^{th}$ comment, and the Evaluation model measures the comment quality to determine whether to end the iterative polishing process or not.
To reduce computation cost and facilitate the sharing of information between models, all $K$ deliberation models share three encoders with others and share the code encoder and comment encoder with the evaluation model.
Each deliberation model has its own decoder, which can avoid these models generating highly similar comments.
% and allow models to iteratively refine the generated comments.
% However, to avoid Deliberation models generate highly similar comments, each Deliberation model have its own decoder
%\lin{this however is strange, cuz I do not see however logic around}. \mfw{update}
We employ a two-step training strategy to train {\tool} as shown in Figure \ref{fig:framework}. In the first step, we locally train the $K$ deliberation models: we first jointly train the first deliberation model and the evaluation model. Then we freeze the shared encoders and train the other deliberation models one by one. 
% After that, we can obtain $K$ well-trained Deliberation models and one well-trained Evaluation model. 
In the second step, we fine-tune {\tool} by jointly optimizing all trained models. 

\subsubsection{Deliberation Model}
Each deliberation model consists of three different encoders (i.e. code encoder, keyword encoder, and comment encoder) and a decoder. 
The details of them are illustrated in 
the following. 
%\lin{I remove Figure 3 here cuz it does not give any new info, please refer to Figure 3 in the following details. I also move the parameter sharing at the end of this section.}
%Figure \ref{fig:delib_model}.

% The Rewrite module is carefully designed to incorporate the context of the past comment and the semantic information of the keywords and the input code to improve the quality of the generated comments. It contains three different encoders and $K$ decoders. The detailed structure is illustrated in Figure \ref{fig:rewrite_module}.

% \begin{figure*}[tbh]
% \centering
% \includegraphics[width=0.95\textwidth]{fig/rewrite_module_v2.pdf}
% \caption{The detailed structure of the Rewrite and the Evaluate module.}
% \label{fig:rewrite_module}
% \end{figure*}

\begin{figure}[tbh]
\centering
\includegraphics[width=0.9\columnwidth]{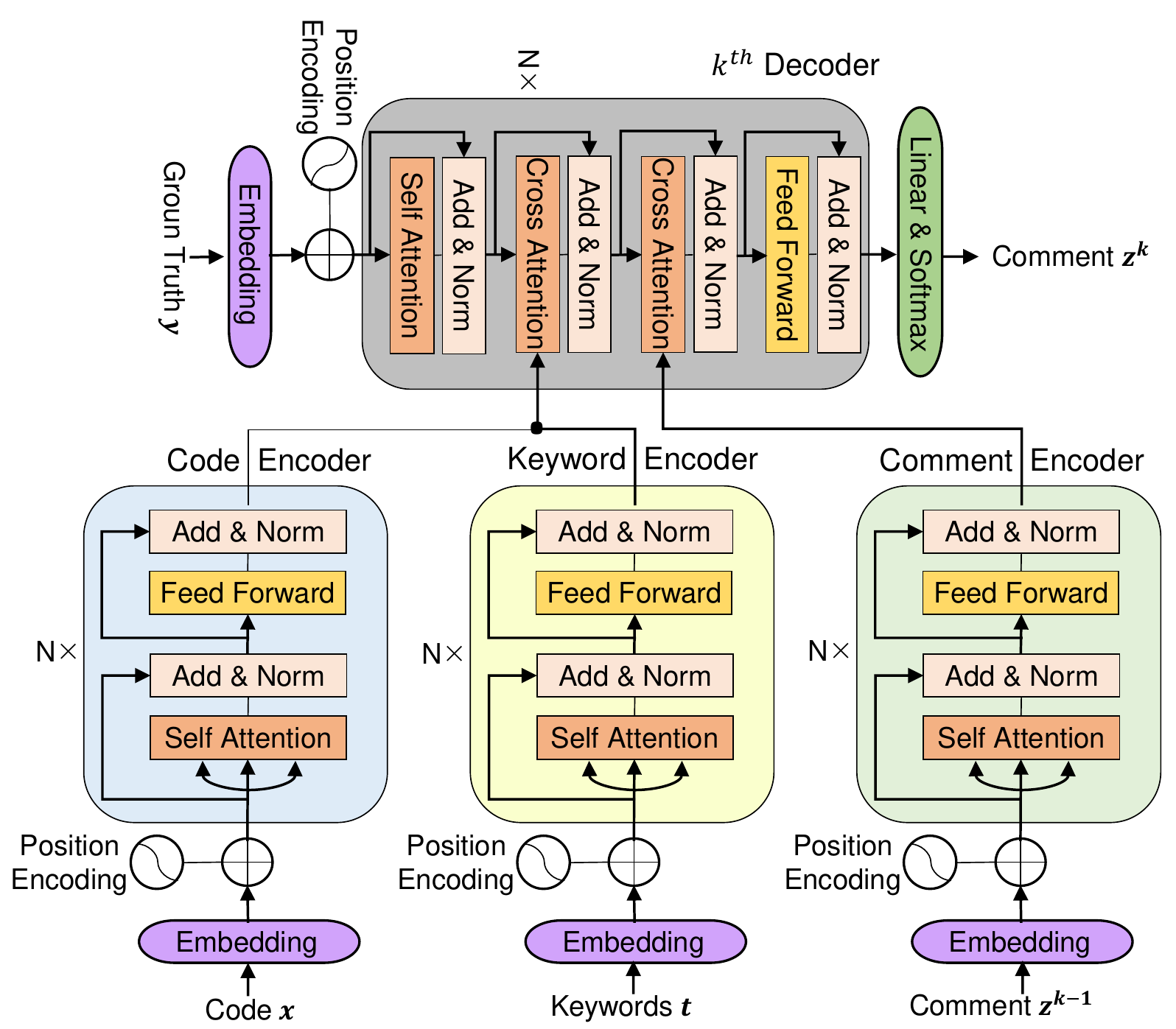}
\caption{The detailed structure of the Deliberation model.}
%\lin{use the same layout with fig 2 if space allowed}}
\label{fig:delib_model}
% \vspace{-0.5cm}
\end{figure}

% \subsubsection{Embeddings}

\textbf{Encoders.}
The code encoder, keyword encoder, and comment encoder aim to encode the source code $x$, keywords $t$, and the previous comment $z^{k-1}$ as vectors, thus enabling the deliberation model to obtain the semantic information from both source-side (code and keywords) and target-side (the past comment).
We construct the three encoders by following the structure of the vanilla Transformer Encoder \cite{DBLP:conf/nips/VaswaniSPUJGKP17}. As shown in Figure \ref{fig:delib_model}, each encoder is composed of a stack of $N$ identical Transformer Encoder blocks. Each block contains two sub-layers: The first sub-layer is a multi-head self-attention layer (MHAtt), which employs multiple attention heads to capture the information from different representation sub-spaces at different positions. The second sub-layer is a two-layer Feed-Forward Network (FFN) with a \texttt{ReLU} activation function in between. The residual connection is employed around the two sublayers, followed by layer normalization (LayerNorm) \cite{DBLP:journals/corr/BaKH16}. Since the three encoders have the same structure, we only introduce the code encoder for simplicity. 
%\mfw{Since the structure of the three encoders is the same as that of , we only introduce the code encoder for simplicity.}
% encode the keywords $T$, the source code $X$, and the past comment $Z^{(k-1)}$ from the $(k-1)$-th turn as vectors, respectively. Note that we use the vanilla Transformer Encoder \cite{DBLP:conf/nips/VaswaniSPUJGKP17} as our encoders, and the three encoders have the same structure but do not share the parameters. Take the code encoder as an example.

{Given a code snippet $x = [x_1, x_2, ..., x_{l(x)}]$, where $l(x)$ is the number of words in the code. The code encoder first embeds each word of the code into a $d$ dimensional word vector:}
\begin{equation}
\overrightarrow{x_i} = {W_e}^\mathrm{\emph{T}}\cdot{x_i} + PE_i \label{eq:embedding}
\end{equation}
where $W_e$ is a trainable embedding matrix, and $PE_i$ is the position encoding of the $i$-th word. Following previous study~\cite{DBLP:conf/nips/VaswaniSPUJGKP17}, we use the $sine$ and $cosine$ function of different frequencies to compute the position encoding:
\begin{align}
    PE_{i,2j} &= \sin{\big(j/{10000}^{2j/d}\big)} \\
    PE_{i,2j+1} &= \cos{\big(j/{10000}^{2j/d}\big)}
\end{align}
where $i$ is the position of the word and $j$ denotes the $j$-th dimension of the embedding vector. 

Then, the code encoder inputs the sequence of word embeddings into $N$ identical encoder blocks to calculate the hidden states of the code. For the $i^{th}$ block of the code encoder, suppose that the input is $H_{i-1}$, the output $H_{i}$ is calculated as follows:
\begin{align}
H_{i,1} &= {\rm LayerNorm}\bigg(H_{i-1} + {\rm MHAtt}(H_{i-1}, H_{i-1}, H_{i-1})\bigg) \label{eq:first_layer} \\
H_{i} &= {\rm LayerNorm}\bigg(H_{i,1} + {\rm FFN} (H_{i,1})\bigg) \label{eq:last_layer}
\end{align}
% \begin{equation}
% H_{i,1} = {\rm LayerNorm}\bigg(H_{i-1} + {\rm MHAtt}(H_{i-1}, H_{i-1}, H_{i-1})\bigg) \label{eq:first_layer}
% \end{equation}
% % Where \texttt{MHAtt} is the Multi-Head Attention \cite{DBLP:conf/nips/VaswaniSPUJGKP17} mechanism, which employs multiple attention heads to capture the information from different representation subspaces at different positions. \texttt{LayerNorm} denotes Layer Normalization operation proposed by Ba et al. \cite{DBLP:journals/corr/BaKH16}.
% The subsequent sub-layer take the $H_{i,1}$ as input, and calculates the output of the $i^{th}$ block $H_{i}$:
% \begin{equation}
% H_{i} = {\rm LayerNorm}\bigg(H_{i,1} + {\rm FFN} (H_{i,1})\bigg) \label{eq:last_layer}
% \end{equation}
% Where \texttt{FFN} is a two-layer Feed-Foward Network with a \texttt{ReLU} activation function in between.
where $H_{i,1}$ is the hidden states of the first sub-layer. Initially, the word embedding vectors $[\overrightarrow{x_1}, \overrightarrow{x_2}, ..., \overrightarrow{x_{l(x)}}]$ are fed into the first block, and the $N^{th}$ block outputs the final hidden states of the input code $H=[h_1, h_2,...,h_{l(x)}]$. Similarly, {\tool} can encode the keywords $t$ and the past comment $z^{k-1}$ into hidden states $P$ and $R^{k-1}$, respectively.

There are two points worth noting: (1) In the first-pass deliberation, {\tool} takes the comment of the retrieved code as the initial draft $z^0$, for each turn after this, {\tool} uses the comment generated in the previous turn as the draft.
(2) source code $x$ and keywords $t$ do not change in the iterative deliberation process, so to save computational resources and time, we compute their hidden states $H$ and $P$ only once, and reuse them in subsequent iterations.

\textbf{Decoder.}
The decoder aims to improve the quality of the previously generated comment $z^{k-1}$ by jointly leveraging its context and the semantics of the source code $x$ and the keywords $t$. 
As shown in Figure \ref{fig:delib_model}, the decoder is also composed of a stack of $N$ identical decoder blocks, and each block consists of four sub-layers. In addition to the first and the last sub-layers {introduced in the part of Encoders in section 2.2.1}, the decoder block inserts two multi-head cross attention sub-layers in between, which are used to capture the information from the outputs of the three encoders.
% Among them, the first and the last sub-layers are the same as those in the encoder we introduced earlier. The two intermediate sub-layers perform cross attention to capture the information output by the three encoders.

In the $k^{th}$ pass deliberation ($k\geq1$), given the hidden states $H$, $P$, $R^{k-1}$. The $i^{th}$ block of the decoder first gets the hidden states of the first sub-layer $S_{i,1}$ using Eq. (\ref{eq:first_layer}). 
Then, in the second sub-layer, the block separately performs multi-head attention over the hidden states of the source code $H$ and the keywords $P$:
\begin{align}
a_i &= {\rm MHAtt} (S_{i,1}, H, H) \\
b_i &= {\rm MHAtt} (S_{i,1}, P, P)
\end{align}
% b_n &= {\rm LayerNorm}\bigg(s_{n_1} + {\rm MHAtt} (s_{n_1}, {h^z}^{(k-1)}, {h^z}^{(k-1)})\bigg) \label{eq:comment_context} \\
% c_n &= {\rm LayerNorm}\bigg(s_{n_1} + {\rm MHAtt} (s_{n_1}, h^t, h^t)\bigg)
Besides, to effectively leverage the information from source-side, we utilize the gate mechanism~\cite{DBLP:journals/neco/HochreiterS97} to adaptively incorporate the $a_i$ containing source code features and the $b_i$ containing keywords features:
\begin{align}
\beta &= {\rm Sigmoid}(W^\mathrm{\emph{T}}_{gate}[a_i~;~b_i]) \\   
S_{i,2} &= {\rm LayerNorm}\bigg(S_{i,1} + \beta\cdot a_i+ (1-\beta)\cdot b_i\bigg)
\end{align}
where $\beta$ is the degree of integration between source code and keywords, A larger value of the $\beta$ (ranges from 0 to 1) indicates that the model should pay more attention to the information in the source code. $W_{gate}$ is a trainable parameter matrix, $[;]$ is concatenation operation, and $S_{i,2}$ is the hidden states of the second sub-layer.
% In particular, based on the Eq. (\ref{eq:comment_context}), the $k$-th decoder can capture the important clues from the contextual information of the past comment for further refinement. 
% Then, the block gets the hidden states of the second sub-layer $s_{n_2}$: 
% \begin{equation}
% s_{n_2} = a_n + b_n + c_n
% \end{equation}
% However, in empirical experiments, we find that low-quality comments may cause the decoder to over-attend noisy tokens, which negatively affects the generation of the new candidate comments. Therefore, we utilize the gate mechanism to adaptively incorporate the $a_n$ containing source code information and the $b_n$ containing past comment information:
% \begin{align}
% &\beta = {\rm Sigmoid}(W^\mathrm{\emph{T}}_{gate}[a_n~;~b_n]) \\   
% &s_{n_2} = \beta\cdot a_n + (1-\beta)\cdot b_n 
% \end{align}
% where $\beta$ is the degree of integration between source code and past comment, A larger value of the $\beta$ (ranges from 0 to 1) indicates that the model should pay more attention to the information in the source code. $W_{gate}$ is a trainable parameter matrix, operation $[;]$ is concatenation, and $s_{n_2}$ is the hidden states of the second sub-layer.
In the third sub-layer, the block obtains the $S_{i,3}$ by performing multi-head attention over the hidden states of the previous comment $R^{k-1}$:
\begin{equation}
S_{i,3} = {\rm LayerNorm}\bigg(S_{i,2} + {\rm MHAtt} (S_{i,2}, R^{k-1}, R^{k-1})\bigg)   
\end{equation}
Based on this equation, the decoder can capture the important clues from the global information of the past comment for further refinement. Then, according to Eq. (\ref{eq:last_layer}), the $i^{th}$ block uses the $S_{i,3}$ to compute the output of the last sub-layer $S_i$. 
After the calculation of $N$ decoder blocks, the decoder gets the hidden states of the last decoder block $S$. For the $j$-th decoding step, the probability of $j^{th}$ token $z_j^k$ can be calculated by projecting the $j^{th}$ state $s_j$ in $S$ via a linear layer followed by a Softmax function.
\begin{equation}
p(z_j^k|z_1^k,~z_2^k,~...,~z^k_{j-1}) = {\rm Softmax}(W_o^\mathrm{\emph{T}}\cdot s_j + b_o) 
\end{equation}
where $W_o$ is the parameter matrix and $b_o$ is the bias. Ultimately, we use the Argmax function to generate the new comment $z^k$.
\begin{equation}
z^k = {\rm Argmax}([p(z_1^k)~;~p(z_2^k)~;~\cdots~;~p(z_{l(k)}^k)])
\end{equation}
where the $l(k)$ is the length of the $k^{th}$ generated comment.

%\song{too much formulas here, hard to follow}

% \textbf{Parameter sharing strategy.} Each Deliberation model shares its three encoders with others\lin{To XXX(reason you design this), all the K Deliberation models share three encoders with others, but have their own decoders}. 
% The decoders of all $K$ Deliberation model have the same structure but do not share the parameters. 

\subsubsection{Evaluation Model}
\begin{figure}[tbh]
\vspace{-0.3cm}
\centering
\includegraphics[width=0.7\columnwidth]{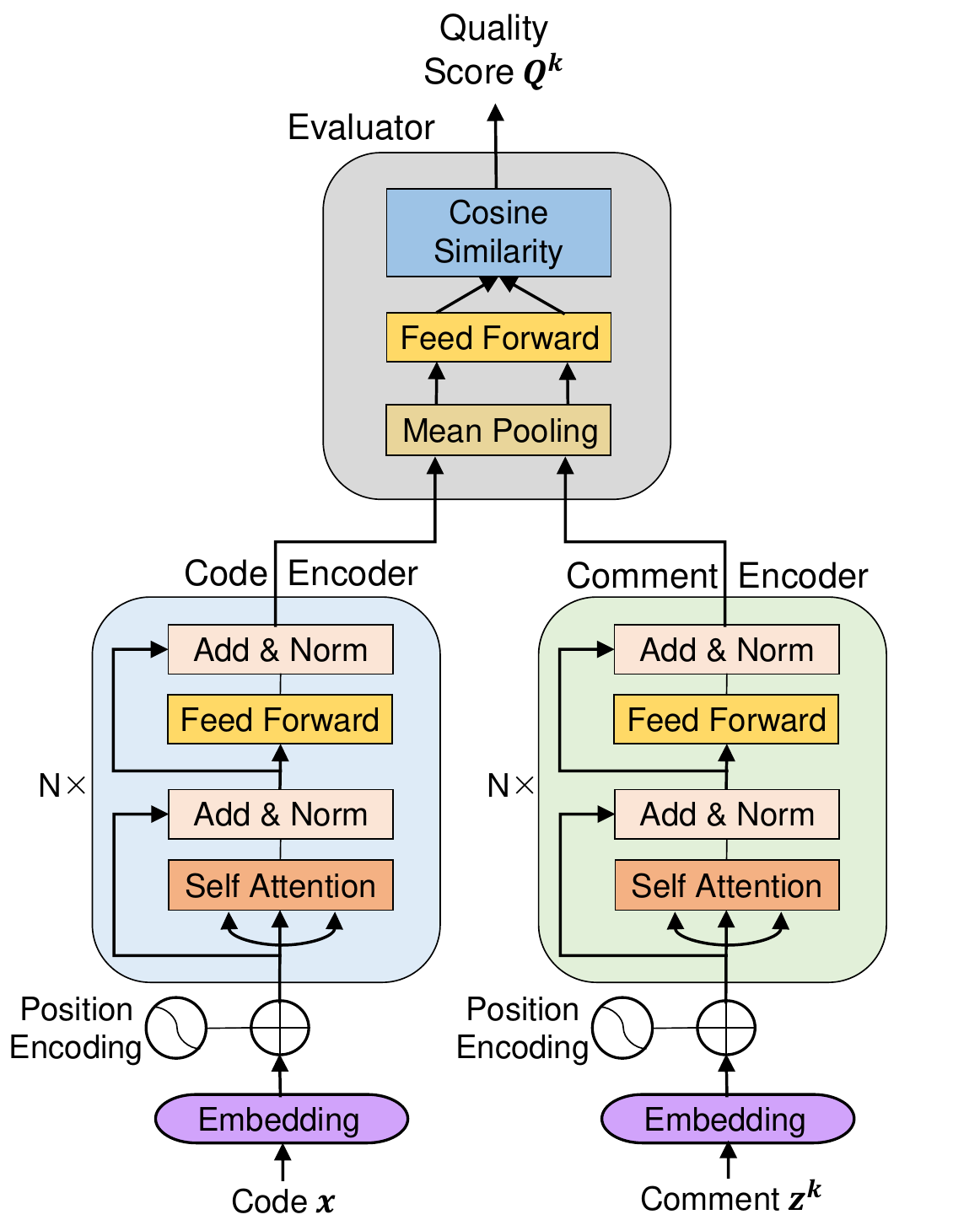}
\caption{The detailed structure of the Evaluation model.}
\vspace{-0.1cm}
\label{fig:eval_model}
\end{figure}
The evaluation model aims to estimate the quality of the generated comments and calculate their quality scores. As shown in Figure \ref{fig:eval_model}, the evaluation model contains a shared code encoder, a shared comment encoder, and an evaluator.
% , where the two encoders are shared with \lin{add some detail}
%\lin{are directly transformed from?}\mfw{``shared with'' may be better? since we need update the parameters of the deliberation model and evaluation model at the same time, ``directly transformed'' feels like we do not train it in evaluation model}
% the Deliberation models.

Given the new comment $z^k$ generated by the $k^{th}$ deliberation model, the comment encoder encodes the $z^k$ into hidden states $R^k$. {To obtain the representation of the comment, the evaluator first uses Mean Pooling to average the hidden states $R^{k}=[r_0^k, r_1^k,...,r_{l(k)}^k]$ to get the aggregated features $v_{mean}^k$. Then, it utilizes a two-layer feed-forward network (FFN) to map the features into the comment representation $v^{k}$}.
\begin{align}
v_{mean}^{k} &= {\rm MeanPooling}([r_0^{k}~;~r_1^{k}~;~\cdots~;~r_{l(k)}^k]) \label{eq:representation_1}\\
v^{k} &= {\rm FFN}(v_{mean}^{k}) = {\rm ReLU}(v_{mean}^{k}\cdot W_1 + b_1)\cdot W_2 + b_2 \label{eq:representation_2}
\end{align}
% where $v_{mean}^{z^{k}}$ is the aggregated features.
Similarly, the evaluator can obtain the code representation $v^x$ using the Eq. (\ref{eq:representation_1}) and (\ref{eq:representation_2}) . Then, we use the cosine similarity metric to calculate the similarity score $Q^{k}$ between $v^{k}$ and $v^x$. A higher similarity score indicates that the comment $z^k$ is more semantically similar to the source code $x$.
\begin{equation}
    Q^{k} = Cos(v^x,~  v^k) = \frac{{v^x}^\mathrm{\emph{T}}\cdot v^k}{\parallel 
    v^x\parallel\cdot \parallel v^k\parallel}
\end{equation}

% \subsection{Model Training}
\subsubsection{Two-Step Training}
% \input{table/alg}
%\song{not sure should we explain model training so complex, just mention parameters to be tuned, and other configurations should be ok}
In this section, we describe the training process and strategies for {\tool}. We denote the parameters of the $k^{th}$ deliberation model as $\theta^k_{d}$ and the parameters of the evaluation model as $\theta_{e}$.
%Assume it is at the $k$-th deliberation, 

\textbf{Deliberation loss and Evaluation loss}.
Given the source code $x$, the ground truth $y$, the keywords $t$, and the previous comment $z^{k-1}$, the $k^{th}$ deliberation model can be optimized by maximizing the probability of $p(y|x,t,z^{k-1})$. The loss function is calculated as:
\begin{equation}
    \mathcal{L}_{delib}(\theta^k_{d}) = \sum\limits_{1\leq i\leq l(y)} -\log p\bigg(y_i|y_{<i},x,t,z^{k-1}\bigg) \label{eq:delib_loss}
\end{equation}
where $l(y)$ is the length of ground truth $y$.
% , $y_{<i}$ is the prior words $[y_0,~y_1,~...,~y_{i-1}]$\lin{what are these prior words?}. 
For the evaluation model, we use the Circle Loss function \cite{DBLP:conf/cvpr/SunCZZZWW20} to optimize its parameters $\theta_{e}$:
\begin{equation}
    \mathcal{L}_{eval}(\theta_{e}) = log\bigg(1 + e^{\lambda(Cos(v^x, v^k)-Cos(v^x, v^y))} \bigg)
\end{equation}
where $Cos()$ denotes the cosine similarity score, $v^x, v^k$ and $v^y$ are the representation vectors of the source code, the $k^{th}$ generated comment, and the ground truth, respectively.
% (\ref{eq:representation_1}) and Eq. (\ref{eq:representation_2}).

% \textbf{Tackling Code Start Problem.}
% In theory we can train the model from random initialization with the following combined loss:
% \begin{equation}
% \begin{aligned}
%     \mathcal{L}(\theta^1_{d},...,\theta^K_{d},\theta_{e})=  & \alpha_1\mathcal{L}_{delib}(\theta^1_{d}) +...+ \alpha_K\mathcal{L}_{delib}(\theta^K_{d})\\ 
%     &+ \alpha_{e}\mathcal{L}_{eval}(\theta_{e}) \label{eq:finetune_loss}
% \end{aligned}
% \end{equation}
% where $\alpha_1, ..., \alpha_K, \alpha_{e}$ are the hyper-parameters which used to control the weights of these losses. However, we find that training the multi-pass model directly from scratch is unstable in practice, which is mainly because the cold start \cite{DBLP:conf/sigir/ScheinPUP02} problem. 
% % Specifically, since the $k^{th}$ Deliberation model depends on the past comment from the $(k-1)$-th turn to generate a new comment, the low-quality comments may cause the Deliberation model to over-attend noisy tokens, which negatively affects the generation of the new comments. Therefore, in the early stage of model training, if the first deliberation model generates a low-quality comment, the new comments generated by subsequent Deliberation models will be impacted, and the negative impact will be propagated.
% \lin{need re-org}

\textbf{Two-step Training Strategy.}
In theory, we can train the framework from random initialization by jointly optimizing all components. However, we find that training the multi-pass model directly from scratch is unstable in practice, which is mainly because of the cold start \cite{DBLP:conf/sigir/ScheinPUP02} problem. To mitigate this problem, we use a two-step training strategy as shown in Figure \ref{fig:framework}.

\textit{Step 1: Locally train the $K$ Deliberation models.} 
We first jointly train the first deliberation model and the evaluation model by minimizing the following loss function:
\begin{equation}
    \mathcal{L}(\theta^1_{d},\theta_{e})= \mathcal{L}_{delib}(\theta^1_{d}) + \alpha_{e}\mathcal{L}_{eval}(\theta_{e})
\end{equation}
where the $\alpha_{e}$ is a hyperparameter, which is set to be 0.1 in our experiments to control the weight of the evaluation loss. Then we freeze the three shared encoders, and iteratively train the subsequent deliberation models using the Eq. (\ref{eq:delib_loss}) until the last deliberation model is trained. 

\textit{Step 2: Globally train the $K$ Deliberation models.} One of the drawbacks of the first-step training is that the deliberation models is optimized independently and the model components cannot share the information. To address this, we further fine-tune {\tool} by jointly training all $K$ deliberation models and the evaluation model:
\begin{equation}
\begin{aligned}
    \mathcal{L}(\theta^1_{d},...,\theta^K_{d},\theta_{e})=  & \mathcal{L}_{delib}(\theta^1_{d}) +...+ \mathcal{L}_{delib}(\theta^K_{d}) +\\ 
    &\alpha_{e}\mathcal{L}_{eval}(\theta_{e}) \label{eq:finetune_loss}
\end{aligned}
\end{equation}
Note that in this step, all parameters are unfrozen and are updated at the same time.

% ``Deep fine-tuning": train the three encoders, the evaluator, and all $K$ decoders jointly with the combined loss $L_{combined}$.
% In theory we can train our approach from random initialization with the following combined loss:
% \begin{equation}
%     L_{combined}= L_{rewrite} + \alpha L_{evaluate}
% \end{equation}
% where $\alpha$ is a hyperparmeter that ranges from 0 to 1.
% However, in practice we find training directly from scratch to be unstable. Therefore, we take a multi-step process to train {\tool}:
% (1).

\subsection{Model Prediction}
The prediction stage aims to generate a concise and useful comment for a given code snippet. As shown in Figure \ref{fig:framework}, given a new code snippet $x$, we first perform data initialization introduced earlier to obtain the keywords $t$ and the initial draft $z^0$. Then, we input them into {\tool} to generate the target comment automatically. The comment generation process involves multiple deliberation processes. During the $k^{th}$ deliberation, the $k^{th}$ deliberation model polishes the previously generated comment $z^{k-1}$ and generates a new comment $z^{k}$. The evaluation model estimates the quality of the new comment $z^{k}$ by calculating the cosine similarity between this and the source code $x$. The deliberation process is performed iteratively unless either of the following two conditions is satisfied: (1) the quality score of the new comment is no longer higher than the previous ones; (2) a certain number of deliberations $K>0$ is reached. In the former case, we adopt the previous comment as the target comment. In the latter case, the last generated comment is accepted.

\section{Experimental Setup}
\label{sec:exp}

\subsection{Dataset} 
Since most of the related studies~\cite{ahmad2020transformer,DBLP:journals/corr/abs-2111-08874,DBLP:journals/corr/abs-2104-09340,DBLP:conf/icse/ZhangW00020,DBLP:journals/corr/abs-2108-00213} for comment generation tasks are evaluated on JCSD \cite{DBLP:conf/ijcai/HuLXLLJ18} and PCSD \cite{barone2017parallel} benchmark datasets, in this study, we also select these two datasets in our experiments.
JCSD has 87,136 code-comment pairs collected from more than 9K Java Github repositories created from 2015 to 2016 with at least 20 stars. 
It first extracted Java methods and Javadocs, and treated the first sentence of the Javadoc as the ground-truth comment of the corresponding code. 
PCSD contains 108,726 code-comment pairs collected from open source repositories on GitHub. It used docstrings (i.e., the string literals that appear right after the definition of functions) as comments for Python functions.

For the sake of fairness, we preprocess the JCSD and PCSD strictly following Rencos \cite{DBLP:conf/icse/ZhangW00020}. Specifically, we first split datasets into a training set, validation set, and test set in a consistent proportion of 8 : 1 : 1 for the Java dataset and 6 : 2 : 2 for the Python dataset. We use the \textit{javalang} \cite{javalang} and \textit{tokenize} \cite{tokenize} libraries to tokenize the code snippet for JCSD and PCSD, respectively. We further split code tokens of the form \emph{CamelCase} and \emph{snake\_case} to respective sub-tokens. 
In common with \cite{DBLP:conf/icse/ZhangW00020}, we remove the exactly duplicated code-comment pairs in the test set for JCSD.
The specific statistics of the two preprocessed datasets are shown in Table \ref{table:dataset}. 

\begin{table}[hbtp!]
\centering
% \vspace{-0.5cm}
\caption{Statistic of Datasets}
\label{table:dataset}
\resizebox{0.69\columnwidth}{!}{
\begin{tabular}{l|c|c}
\hline
Dataset                  & JCSD    & PCSD  \\
\hline
Train                    & 69,708  & 65,236  \\
Validation               & 8,714   & 21,745  \\
Test                     & 6,489   & 21,745  \\
\hline
Unique tokens in code    & 230,336 & 481,756 \\
Unique tokens in comment & 35,535  & 37,111  \\
\hline
Avg. tokens in code      & 99.9    & 133.1   \\
Avg. tokens in comment   & 17.1    & 9.9     \\
\hline
Max. token in code       & 4,842   & 157,116 \\
Max. token in comment    & 670     & 333    \\
\hline
\end{tabular}
\vspace{-0.6cm}
}
\end{table}

\subsection{Evaluation Metrics}

We evaluate the performance of different approaches using common metrics including BLEU~\cite{DBLP:conf/acl/PapineniRWZ02}, ROUGE-L~\cite{lin2004rouge}, METEOR~\cite{DBLP:conf/acl/BanerjeeL05}, and CIDEr~\cite{DBLP:conf/cvpr/VedantamZP15}.
\textbf{BLEU} measures the $n$-gram precision by computing the overlap ratios of $n$-grams and applying brevity penalty on short translation hypotheses. 
BLEU-1/2/3/4 corresponds to the scores of unigram, 2-grams, 3-grams, and 4-grams, respectively. 
\textbf{ROUGE-L} is defined as the length of the longest common subsequence between generated sentence and reference, and based on recall scores.
\textbf{METEOR} is based on the harmonic mean of unigram precision and recall, with recall weighted higher than precision. 
\textbf{CIDEr} considers the frequency of $n$-grams in the reference sentences by computing the TF-IDF weighting for each $n$-gram. ${CIDEr}_n$ score for $n$-gram is computed using the average cosine similarity between the candidate sentence and the reference. 

%To ensure fairness, we compute the values of these automated metrics following the same scripts used by Rencos \cite{}.

\subsection{Implementation Details}

Following previous studies \cite{DBLP:conf/icse/ZhangW00020}, we set the length limits (in terms of \#words) of code and comment (i.e., 300 and 30 for JCSD, 100 and 50 for PCSD). 
To save the computing resource, we limit the maximum vocabulary size of source code and comment to 50K for both datasets. The out-of-vocabulary words are replaced by `UNK'. 
The word embedding size of both code and comment is set to 512. We set the dimensions of hidden states to 512, the number of heads to 8, and the number of blocks to 6, respectively. The maximum deliberation number $K$ is set to be 3. We set the mini-batch size to 32 and train our approach using the Adam \cite{DBLP:journals/corr/KingmaB14} optimizer. In the first-step training, we set the learning rate to 1e-4, and for the second-step training, {we use a smaller learning rate (1e-5) to fine-tune {\tool}}. To avoid the over-fitting problem, we apply dropout \cite{DBLP:journals/corr/abs-1207-0580} with 0.2. The maximum number of epochs is set to 100 for each step of training. We also use the strategy of early stopping, {when} the validation performance does not improve for 20 consecutive epochs, the training process will be stopped. To reduce training time, we use the greedy search to generate comments at the training stage. During the prediction stage, we use the beam search~\cite{DBLP:conf/emnlp/WisemanR16} and set the beam size to 5 for choosing the best result. Our approach is implemented based on the Pytorch \cite{pytorch} framework. The experimental environment is a desktop computer equipped with an NVIDIA GeForce RTX 3060 GPU, intel core i5 CPU, and 12GB RAM, running on Ubuntu OS.

\section{Results}
\label{sec:result}
We address the following three research questions to evaluate the performance of {\tool}:

\textbf{RQ1 }: How does the {\tool} perform compared to the state-of-the-art comment generation baselines?

\textbf{RQ2}: How does each individual component in {\tool} contribute to the overall performance?

\textbf{RQ3}: What's the performance of {\tool} on the data with different code or comment length?

% 勿删，这个表格里CIDEr保留三位小数。
\begin{table*}[t]
\centering
\caption{The results of comparison with baselines, with the improvement compared with the best baselines in percentage.}
\label{table:RQ1_results}
\vspace{-0.1in}
\resizebox{\textwidth}{!}{
\begin{tabular}{l|ccccccc|ccccccc}
% \toprule
\hline
\multicolumn{1}{c|}{\multirow{2}{*}{\textbf{Method}}} & \multicolumn{7}{c|}{JCSD}                                    & \multicolumn{7}{c}{PCSD}                                    \\
% \cmidrule(lr){2-8}\cmidrule(lr){9-15}
\cline{2-15}
\multicolumn{1}{c|}{}                        & \multicolumn{4}{c}{BLEU-1/2/3/4} & ROUGE-L & METEOR & CIDEr & \multicolumn{4}{c}{BLEU-1/2/3/4} & ROUGE-L & METEOR & CIDEr \\
% \midrule
\hline
LSI                                         & 31.4   & 22.5   & 19.3   & 17.3  & 34.8    & 14.4   & 1.803 & 36.3   & 23.6   & 20.1   & 17.6  & 40.0    & 17.2   & 1.982 \\
VSM                                         & 33.3   & 24.4   & 21.1   & 19.0  & 36.6    & 15.4   & 1.983 & 38.9   & 26.1   & 22.1   & 19.3  & 42.7    & 19.0   & 2.216 \\
NNGen                                       & 33.0   & 24.4   & 20.9   & 18.7  & 36.3    & 15.0   & 1.933 & 36.5   & 23.8   & 20.1   & 17.4  & 40.2    & 17.1   & 1.967 \\
% \midrule
\hline
CODE-NN                                     & 23.9   & 12.8   & 8.6    & 6.3   & 28.9    & 9.1    & 0.978 & 30.8   & 15.4   & 10.7   & 8.1   & 35.1    & 13.4   & 1.229 \\
TL-CodeSum                                  & 29.9   & 21.3   & 18.1   & 16.1  & 33.2    & 13.7   & 1.660 & 31.1   & 16.5   & 12.5   & 10.4  & 35.3    & 13.6   & 1.335 \\
Hybrid-DRL                                  & 32.4   & 22.6   & 16.3   & 13.3  & 26.5    & 13.5   & 1.656 & 41.1   & 26.2   & 19.5   & 15.0  & 42.2    & 17.9   & 2.042 \\
% \midrule
\hline
Re$^2$com                                      & 33.7       & 23.6       & 19.0       & 16.3      & 38.1        & 15.1       & 1.807      & 36.6       & 22.3       & 17.4       & 14.5      & 40.8        & 17.0       & 1.813      \\
Rencos                                      & 37.5   & 27.9   & 23.4   & 20.6  & 42.0    & 17.3   & 2.209 & 43.1   & 29.5   & 24.2   & 20.7  & 47.5    & 21.1   & 2.449 \\
EditSum                                     & 34.1       & 24.3       & 19.5       & 16.9      &    38.6     & 15.2       & 1.865      &    37.7    & 23.1       & 18.2       & 15.6     & 42.0        &   17.1     &  1.894     \\
\hline
\textit{\tool}                                & \textbf{40.4}   & \textbf{30.2}  & \textbf{25.2}  &  \textbf{22.3}  &  \textbf{44.5}    &  \textbf{19.6}  &  \textbf{2.442} &  \textbf{45.6} &  \textbf{31.4}  &  \textbf{25.5}  &  \textbf{21.9} &  \textbf{49.3}   &  \textbf{22.5}  &  \textbf{2.603}\\
% \bottomrule
\hline
\end{tabular}
}
\end{table*}
% \usepackage{multirow}
\begin{table*}[t]
\centering
\caption{RQ2 Ablation study on the multi-pass deliberation and evaluation model.}
\label{table:RQ2_results}
\vspace{-0.1in}
\resizebox{\textwidth}{!}{
\begin{tabular}{c|ccccccc|ccccccc}
% \toprule
\hline
\multicolumn{1}{c|}{\multirow{2}{*}{Variants}}                        & \multicolumn{7}{c}{JCSD}                                                                                       & \multicolumn{7}{|c}{PCSD}                                                                                       \\
% \cmidrule(lr){2-8}\cmidrule(lr){9-15}
\cline{2-15}
                                                 & \multicolumn{4}{c}{BLEU-1/2/3/4}                              & ROUGE-L       & METEOR        & CIDEr          & \multicolumn{4}{c}{BLEU-1/2/3/4}                              & ROUGE-L       & METEOR        & CIDEr          \\
                                                %  \midrule
                                                \hline
{\tool} w/o Multi-pass Deliberation & 38.9           & 28.5          & 23.5          & 20.8          & 43.1          & 18.8          & 2.274          & 43.5          & 29.3          & 23.8          & 20.4          & 47.5          & 21.1          & 2.424          \\
{\tool} w/o Evaluation Model           & 39.5          & 29.3          & 24.3          & 21.5          & 43.7          & 19.0          & 2.338          & 44.6          & 30.3          & 24.3          & 20.6          & 48.6          & 21.6          & 2.478          \\
% {\tool} w/o Keywords Encoder      & 39.9          & 29.8          & 24.8          & 22.0          & 44.2          & 19.4          & 2.392          & 45.1          & 31.0          & 25.2          & 21.4          & 48.9          & 22.2          & 2.542          \\
% {\tool} w/o Globally Finetune      & 39.9          & 29.8          & 24.8          & 22.0          & 44.2          & 19.4          & 2.392          & 44.3          & 30.2         & 24.2          & 20.5          & 48.1          & 21.5          & 2.462          \\
{\tool}                         & \textbf{40.4} & \textbf{30.2} & \textbf{25.2} & \textbf{22.3} & \textbf{44.5} & \textbf{19.6} & \textbf{2.442} & \textbf{45.6} & \textbf{31.4} & \textbf{25.5} & \textbf{21.9} & \textbf{49.3} & \textbf{22.5} & \textbf{2.603}\\
% \bottomrule
\hline
\end{tabular}
}
\end{table*}
\subsection{RQ1: Comparison with Baselines}

\subsubsection{Baselines.}
We compare our approach with three categories of existing work on the comment generation task. We exactly adopt the hyperparameter settings reported in the original paper for all baselines. 
For a fair comparison, we use the same maximum code and comment length for all approaches, and evaluate their performance using the same training/testing datasets.

\begin{itemize}[leftmargin=*]
\item{IR-based baselines.}
 \textbf{LSI \cite{DBLP:journals/jasis/DeerwesterDLFH90}} is an IR technique to analyze the latent meaning or concepts of documents. 
 The similarity {between the code and the comment} is computed based on the LSI-reduced vectors and cosine distance, and we set the vector dimension to be 500.
\textbf{VSM \cite{DBLP:journals/cacm/SaltonWY75}} is also a commonly used IR technique in comment generation tasks. For a given code snippet, we represent the code as a vector using TF-IDF, and extract the comment of the most similar code based on cosine similarity.
\textbf{NNGen~\cite{liu2018neural}} is a nearest-neighbors approach for generating commit messages. It first embeds code into vectors based on the bag of words and the term frequency. Then, it retrieves the nearest neighbors of the code. Finally, it outputs the message of the code with the highest BLEU score.

\item{NMT-based approaches.}
%\begin{itemize}[leftmargin=4mm]
\textbf{CODE-NN \cite{DBLP:conf/acl/IyerKCZ16}} is the first learning-based model for comment generation. It maps the source code sequence into word embeddings, then uses the LSTM and the attention mechanism to generate comments.
\textbf{TL-CodeSum \cite{DBLP:conf/ijcai/HuLXLLJ18}} is a multi-encoder neural model that encodes API sequences along with code token sequences and generates comments from source code with transferred API knowledge. 
\textbf{Hybrid-DRL \cite{DBLP:conf/kbse/WanZYXY0Y18}} incorporates ASTs and sequential content of code snippets into a deep reinforcement learning framework.

\item{Hybrid approaches.}
%\begin{itemize}[leftmargin=4mm]
\textbf{Rencos \cite{DBLP:conf/icse/ZhangW00020}} is a hybrid approach that combines the advantages of both IR-based and NMT-based techniques. 
\textbf{Re$^2$Com \cite{DBLP:conf/kbse/WeiLLXJ20}} is an exemplar-based comment generation approach that leverages the advantages of three types of methods based on neural networks, templates, and IR to improve the performance. 
\textbf{EditSum \cite{DBLP:conf/kbse/LiL000J21}} is the most recent hybrid approach.
It first retrieves the most similar code snippet, and treats the corresponding comment as a prototype. Then, it combines the pattern in the prototype and semantic information of the input code to generate the target comment.
 
 \end{itemize}

\subsubsection{Results.}
Table \ref{table:RQ1_results} shows the comparison results between the performance of {\tool} and other baselines, and the best performance is highlighted in bold. 
Overall, our approach achieves the best performance on all evaluation metrics, followed by Rencos, EditSum, and Re$^2$com. 
On the Java dataset, {{\tool} achieves 22.3, 44.5, 19.6, and 2.442 points on BLEU-4, ROUGE-L, METEOR, and CIDEr.} 
Compared with the best baseline (Rencos), {\tool} improves the performance of BLEU-4, ROUGE-L, METEOR, and CIDEr by 8.3\%, 6.0\%, 13.3\%, and 10.5\%, respectively. 
On the Python dataset, {{\tool} achieves 21.9, 49.3, 22.5, and 2.603 points on BLEU-4, ROUGE-L, METEOR, and CIDEr.} 
Compared with the best baseline (Rencos), {\tool} also achieves 5.8\%, 3.8\%, 6.6\%, and 6.3\% improvements on BLEU-4, ROUGE-L, METEOR, and CIDEr, respectively .

%\begin{tcolorbox}[width=8.5cm]
\textbf{Answering RQ1:} {\tool} outperforms the state-of-the-art baselines in terms of all seven metrics on both two datasets. Compared to the best baseline Rencos, {\tool} improves the performance of BLEU-4, ROUGE-L, METEOR, and CIDEr by 8.3\%, 6.0\%, 13.3\%, and 10.5\% on JCSD dataset, by 5.8\%, 3.8\%, 6.6\%, and 6.3\% on PCSD dataset, respectively.
%\end{tcolorbox}

\subsection{RQ2: Component Analysis}

\subsubsection{Variants.}
To evaluate the contribution of core components, we obtain two variants: (1) \textbf{{\tool} w/o Multi-pass Deliberation}, which removes the multi-pass deliberation and adopts the one-pass process to generate comments. (2) \textbf{{\tool} w/o Evaluation Model}, {which removes the evaluation model and takes the comment generated by the last ($K^{th}$) deliberation model as the result}. 
{We train the two variants with the same experimental setup as {\tool} and evaluate their performance on the test sets of JCSD and PCSD, respectively.}

\subsubsection{Results.}
Table \ref{table:RQ2_results} presents the performances of {\tool} and its two variants. 
We can see that, removing the two components makes the performance degrade substantially. Specifically, when comparing {\tool} and {\tool} w/o Multi-pass Deliberation, removing the multi-pass deliberation will lead to a dramatic decrease in the average BLEU-4 (by 6.8\%), ROUGE-L (by 3.4\%), METEOR (by 5.2\%), and CIDEr (by 6.9\%) across both datasets. 
{When comparing {\tool} and {\tool} w/o Evaluation Model,
we find that removing the evaluation model will lead to the performance decline in the average BLEU-4 (by 4.8\%), ROUGE-L (by 1.6\%), METEOR (by 3.5\%), and CIDEr (by 4.5\%).}
We can also observe that, removing the multi-pass deliberation will lead to a larger degree of performance decline than removing the evaluation model. 
%Thus, the multi-pass deliberation contributes more to the performance of {\tool}.
%This indicates that the multi-pass deliberation is an essential component in contributing to {\tool}’s high performances. 

% \song{As the difference between these two variants is marginal in values, we need a statistical test to show the difference are significant}\lin{Mann U shows not significant}

% It is mainly because the evaluation model can not only help {\tool} to terminate the iterative deliberation process and choose the best-generated comment, but also learn a better representation for code and comment by sharing its encoders with the deliberation models.
%We also note that the removal of multi-pass deliberation leads to more performance decline than the removal of the evaluation model, which indicates that the multi-pass deliberation contributes more to the performance of {\tool}. 

% This indicates that the multi-pass decoding is an essential component to contribute to {\tool}’s high performances.
% When compared with {\tool} and {\tool} w/o keyword information, removing the keyword information well lead to the average decrease of BLEU-4 (by 1.8\%), ROUGE-L (by 0.7\%), METEOR (by 1.2\%), CIDEr (by 2.2\%).

%\begin{tcolorbox}[width=8.5cm]
\textbf{Answering RQ2:} Both the multi-pass deliberation and the evaluation model components have  positive contributions to the performance of {\tool}, where the multi-pass deliberation component contributes more to increasing the performance.
%\end{tcolorbox}
%\subsection{RQ3: Tackling Low-Frequency Keywords}

\subsection{RQ3: Performance for Different Lengths}

\begin{figure*}[t]
\centering
\includegraphics[width=\textwidth]{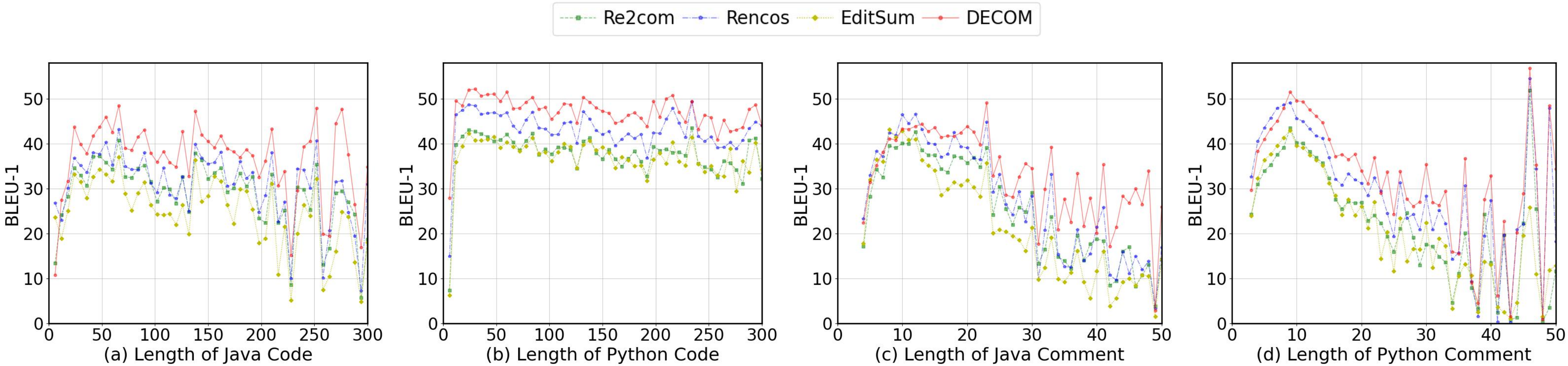}
\vspace{-0.2in}
\caption{BLEU-1 scores for different code and comment lengths.}%\song{only show BLUE-1?}
\label{fig:length}
\end{figure*}

\begin{figure*}[h]
\centering
\includegraphics[width=\textwidth]{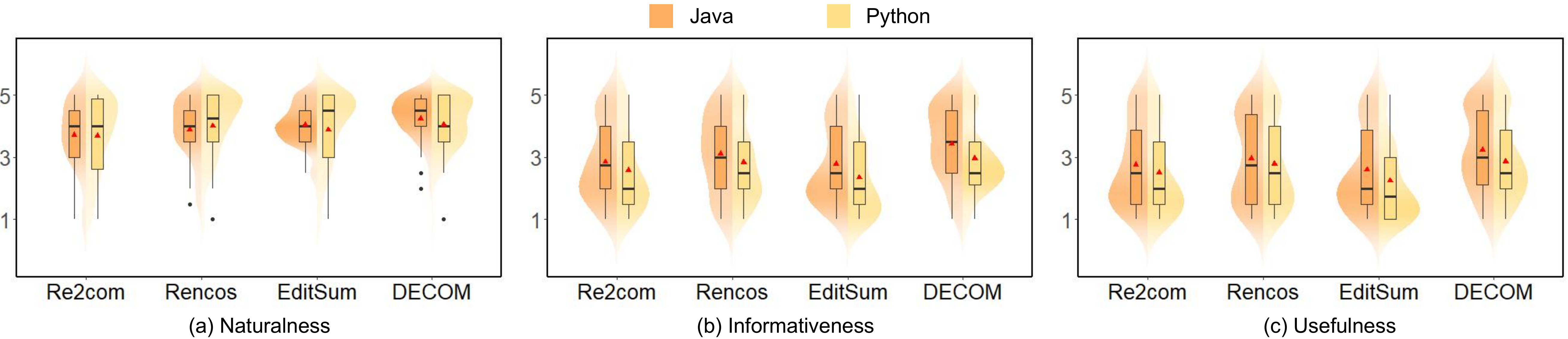}
\vspace{-0.2in}
\caption{The results of human evaluation.}
\vspace{-0.1in}
\label{fig:HE_all}
\end{figure*}

\subsubsection{Methodology}
{To answer this question, we analyze the performance of {\tool} and {best three baselines (i.e. Re$^2$com, Rencos, and EditSum)} 
%\song{give 1-2 sentence to justify when only select 3 here}
on different lengths (i.e., number of tokens) of code and comments.}
We calculate the BLEU-1 score of each sample on the test set of both datasets and average the scores by the length of code and comments, respectively. {(Note that, based on our observations, all the seven evaluation metrics show similar trends. For simplicity, we show BLEU-1 only).} 
% We choose different length ranges for code and comments on the two datasets based on the experimental setup of the methods. For example, on the JCSD dataset, the maximum length of comments is limited to 30. Therefore, we analyze the performance of these methods on JCSD with the comment length ranges from 0 to 30. 

\subsubsection{Results}
Figure \ref{fig:length} presents the performance of {\tool} and the three baselines on JCSD and PCSD datasets with code and comments of different lengths, where the red lines denote the performance of {\tool}.
Overall, we can observe that 
the performance of {\tool} generally outperforms the three baselines with different code and comment lengths on both datasets.
Specifically, as the length of the input code increases, {\tool} almost keeps a stable improvement over the other three approaches. %\song{seems when lengths <5, {\tool} cannot outperform others}
The performance of {\tool} is nearly the best on all the lengths of Java and Python
%\song{Java, Python, globally fix} 
code snippets. 
In particular, {\tool} can achieve much higher performance than others when the length of the Java code snippet is over 200 words. This shows that {\tool} can better understand the semantics of the long code snippets by sharing the information between deliberation models and the evaluation model.
For the output comments, we can see that when the output comments are becoming complicated with a relatively long length, the performance of all the approaches decrease, which indicates that the longer the comment, the harder to generate it completely. However, {\tool} still has a substantial improvement over the other baselines (as shown in Figure 5(c)), showing that our approach has the ability to generate long and concise comments.

\textbf{Answering RQ3:} {\tool} generally outperforms the {best} three baselines on different lengths of the input code snippets and the output comments, indicating its robustness. 
In particular for Java, {\tool} can achieve much higher performance than others when the code snippets and comments are long.
%\end{tcolorbox}

\section{Human Evaluation}
Although the evaluation metrics (i.e., BLEU, ROUGE-L, METEOR, and CIDEr) can measure the lexical gap between the generated comments and the references,
it can hardly reflect the semantic gap. 
Therefore, we perform a human evaluation to further assess the quality of comments generated by different approaches. 

% \textbf{Human Annotators}. {We recruit six participants, including three PhD students, one master student, and two senior researchers. They all have at least two years of software development experience, and four of them have more than five years of development experience.}

\subsection{Procedure}
We recruited six participants, including three Ph.D. students, one master student, and two senior researchers, who are not co-authors of this paper. They all have at least three years of both Java and Python development experience, and four of them have more than six years of development experience. 
We randomly select 100 code snippets from the test dataset (50 from JCSD and 50 from PCSD). By applying the best three baselines (i.e., Re$^2$com, Rencos, and EditSum) and {\tool}, we obtain a total of 400 generated comments. The 400 code-comment pairs are divided into three groups, and each group is used to create a questionnaire. We randomly list the code-comment pairs on the questionnaire and remove their labels to ensure that the participants are not aware of where the comments are generated from. Each questionnaire is evaluated by two participants, and the final result of a generated comment is the average of two participants. 
Each participant is asked to rate each generated comment from the three aspects: (1) \textbf{Naturalness} reflects the fluency of generated comments from the perspective of grammar; (2) \textbf{Informativeness} reflects the information richness of generated comments; and (3) \textbf{Usefulness} reflects how can generated comments help developers. All three scores are integers, ranging from 1 to 5 (1 for poor, 2 for marginal, 3 for acceptable, 4 for good, and 5 for excellent). 
% {Higher score means more positive.} 
%Each group is evaluated by two participants, and the score of a sample is the average of two participants. 

%In Section 4.3, we use BLEU, ROUGE-L, METEOR and CIDEr to compare the performance of  models. Unfortunately, the above metrics can only measure the lexical similarity between the reference summaries and generated results\lin{change the negative tone, following }. In order to further compare the semantic similarity between them, we conducted a manual evaluation, considering three metrics: naturalness, informativeness and usefulness\cite{DBLP:conf/kbse/LiL000J21}.

% \begin{figure*}[thbp]
% \centering
% \subfigure[Naturalness]{
%         \label{fig:HE_1}
%         \includegraphics[width=0.67\columnwidth,height=3.2cm]{fig/HE_1.pdf}
    
% }
% \subfigure[Informativeness]{
%         \label{fig:HE_2}
%         \includegraphics[width=0.67\columnwidth,height=3.2cm]{fig/HE_2.pdf}
    
% }
% \subfigure[Usefulness]{
%         \label{fig:HE_3}
%         \includegraphics[width=0.67\columnwidth,height=3.2cm]{fig/HE_3.pdf}
% }
% \vspace{-0.5cm}
% \caption{The results of human evaluation\cx{add legend}}
% \label{fig:HE}
% \end{figure*}
\subsection{Results}
\begin{table}[htp]
\centering
% \vspace{-0.1in}
\caption{The statistic results of human evaluation.}
% \vspace{-0.1in}
\label{table:HE}
\resizebox{0.75\columnwidth}{!}{
\begin{tabular}{c|c|c|c|c}
\hline
\multicolumn{1}{c|}{}  & Approach   & Avg.         & Median & Std. \\
\hline
\multirow{4}{*}{Naturalness}     & Re$^2$com   & 3.7          & 4.0    & 1.1  \\
                                 & Rencos  & 3.9          & 4.0    & 1.0  \\
                                 & EditSum & 4.0          & 4.0    & 0.9  \\
                                 & DECOM   & \textbf{4.1} & 4.5    & 0.8  \\
                                 \hline
\multirow{4}{*}{Informativeness} & Re$^2$com   & 2.7          & 2.5    & 1.3  \\
                                 & Rencos  & 3.0          & 3.0    & 1.3  \\
                                 & EditSum & 2.6          & 2.0    & 1.3  \\
                                 & DECOM   & \textbf{3.2}          & 3.0    & 1.2  \\
                                 \hline
\multirow{4}{*}{Usefulness}      & Re$^2$com   & 2.6          & 2.0    & 1.4  \\
                                 & Rencos  & 2.9          & 2.5    & 1.4  \\
                                 & EditSum & 2.4          & 2.0    & 1.3  \\
                                 & DECOM   & \textbf{3.1} & 3.0    & 1.3 \\
                                 \hline
\end{tabular}
\vspace{-0.1in}
}
\end{table}
Figure \ref{fig:HE_all} exhibits the results of human evaluation by showing the violin plots depicting the naturalness, informativeness, and usefulness of different models, and Table~\ref{table:HE} shows the statistic results. Each violin plot contains two parts, i.e., the left and right parts reflect the evaluation results of models on the JCSD dataset and PCSD dataset. The box plots in the violin plots present the distribution of data and the red triangles mean the average scores of the three aspects. 
Overall, {\tool} is better than all baselines in three aspects. 
The average score for naturalness, informativeness, and usefulness of our approach are 4.24, 3.43, and 3.25, respectively, on the JCSD dataset. On the PCSD dataset, our approach gets the average score of 4.05, 2.96, and 2.87 in terms of naturalness, informativeness, and usefulness.
{We can see that, the comments generated in the PCSD dataset receive lower scores in human evaluation, while receiving higher scores in evaluation metrics (see Table~\ref{table:RQ1_results}). 
This is mainly because the PCSD dataset contains shorter comments (see Table~\ref{table:dataset}), thus mistakenly generating fewer keywords may lead to a lower degree of human satisfaction. While the shorter comments are more probable with these N-gram matching metrics~\cite{reiter2018structured}.}
%\song{1-2 sentence to explain why on JCSD is better}
%\lin{no idea, some comparison info: 1.PCSD has longer code but shorter comment see table 1; 2. PCSD achieves better performance in table 2; 3. table 4 shows Python comment has one error. My guess: does it because pcsd has shorter comments so are likely to have larger BLEU, but some keywords are hard to predict, and missing these keyword will make readers more unsatisfactory?}for the N-gram matching metrics, such as BLEU, are more probable with shorter texts.

Specifically, in terms of naturalness, our approach achieves average scores above 4 on both JCSD and PCSD datasets, which shows that {\tool} can generate fluent and readable comments. Besides, in terms of informativeness and usefulness, {\tool} is the only approach with an average score of more than 3 points on the JCSD dataset. It indicates that the comments generated by {\tool} tend to be more informative and useful than other baselines. 
%\song{if possible use a table to show the average, sd, median of the four approaches in each aspect; as the figures cannot distinguish the difference well, results look similar}\cx{chenxiao, please make one first see if we have space}

% For the naturalness aspect, the average score of our approach is at least XX\%, XX\%, and XX\% higher than that of Re$^2$com, Rencos, and EditSum on both datasets, respectively. For the informativeness aspect, the average score of our approach is at least XX\%, XX\%, and XX\% higher than that of the three baselins. For the usefulness aspect, the average score of our approach is at least XX\% higher than that of the best baseline XXX.
% The values of usefulness are lower than the corresponding values of informativeness for all approaches. The possible reason is that the generated summary contains not only the information of the code, but also other redundant information. 

\section{Discussion}
\label{sec:discussion}
\subsection{Qualitative Analysis}

\begin{figure}[t]
\centering
\includegraphics[width=\columnwidth]{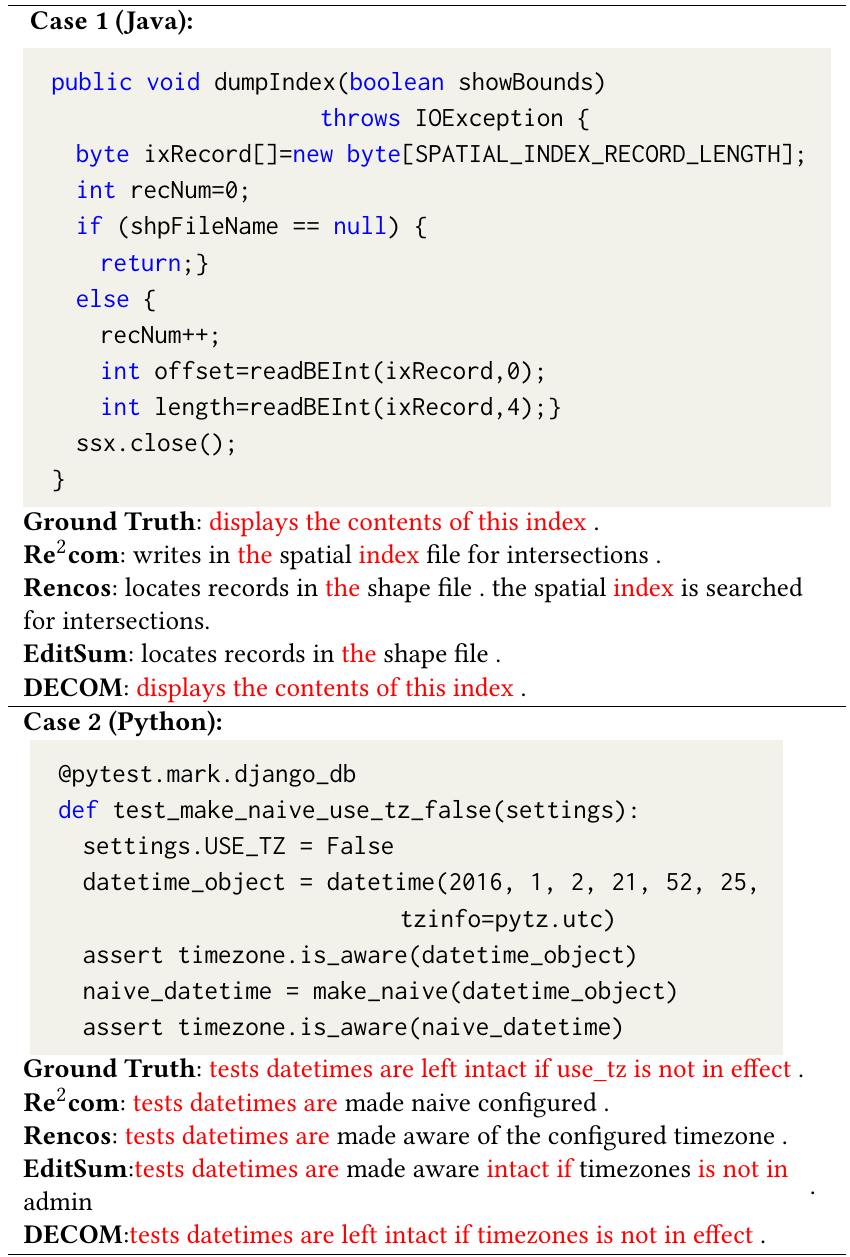}
\vspace{-0.1in}
\caption{Examples of qualitative analysis.}
\label{fig:example}
\end{figure}

For qualitative analysis of our approach, we present two cases generated by the best three baselines together with {\tool}. The cases are selected from the test sets of Java and Python datasets respectively, as shown in Figure \ref{fig:example} .
Overall, the comments generated by {\tool} tend to be more accurate and more readable than the other three baselines.
In case 1, the aim of the Java code is to display the contents of an index. The three baselines mistakenly predict the keyword ``displays'' as ``locates'', ``writes'', and ``locates'', respectively, {resulting in the semantics of the generated comments being different from the ground truth}. In contrast, the comment generated by {\tool} is exactly the same as the ground truth, indicating that our approach can understand the intention of code concisely.
In case 2, we can see that, our approach also performs better than other baselines, and the comment generated by {\tool} has a high semantic similarity with the ground truth.

We believe that the performance advantage of {\tool} mainly comes from two aspects: (1) {\tool} can observe the entire previously generated comment and leverage its global information to polish it. While other baselines can only leverage the previously generated words.
(2) {\tool} employs an evaluation model that can determine the opportunity when the deliberation process should end, as well as learn the semantic relationship between source code and target comments. Besides, the evaluation model shares its two encoders with the deliberation models, which facilitates the information sharing among these models, and enables {\tool} to learn a better representation for code and comments.

\subsection{Parameter Analysis}

\begin{figure}[h]
\centering
\includegraphics[width=0.6\columnwidth]{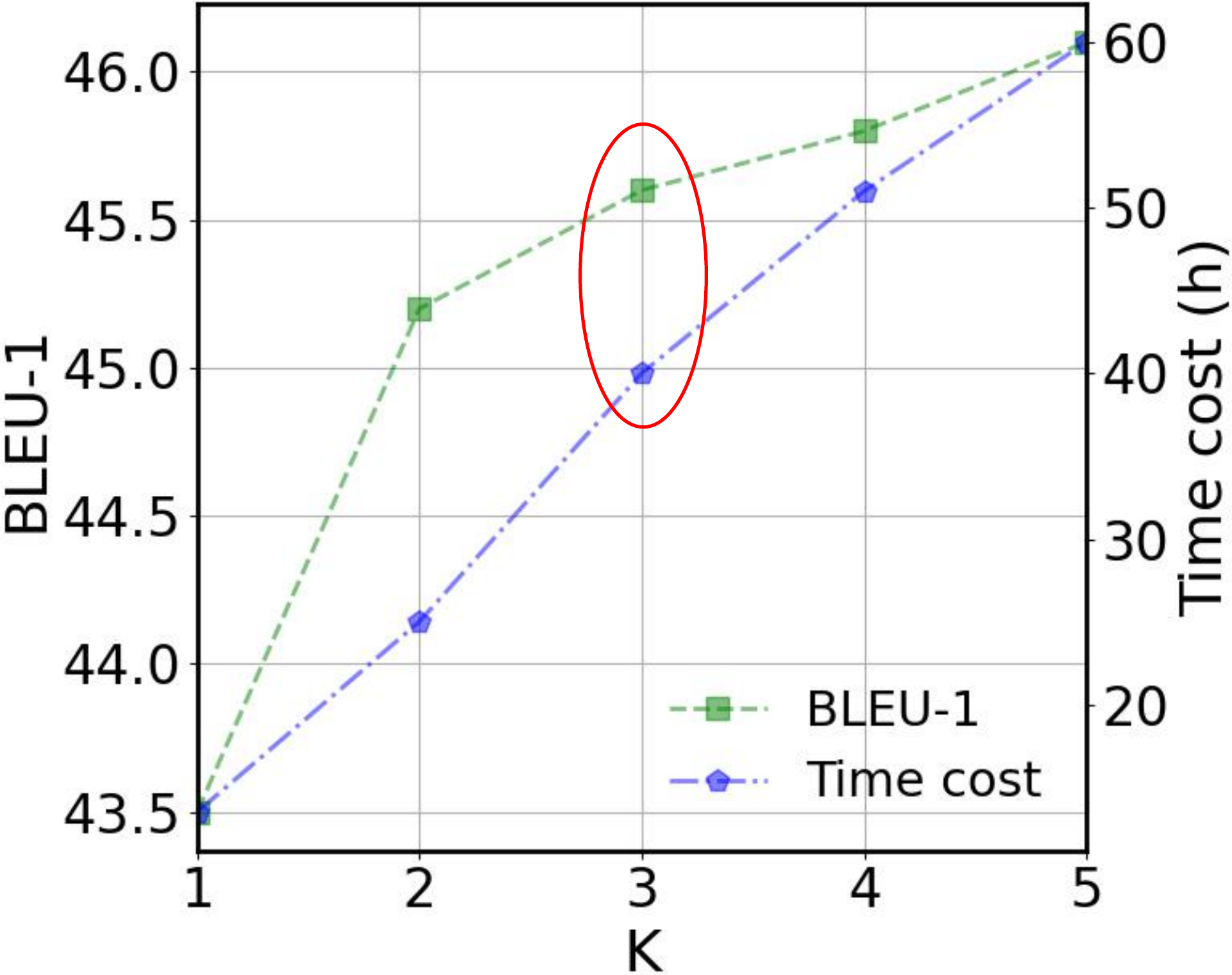}
\caption{Performance and time cost in varying the maximum number of deliberations $K$.}
\vspace{-0.1in}
\label{fig:sensitive}
\end{figure}

Figure \ref{fig:sensitive} illustrates the impact of the maximum number of deliberations $K$ on the performance of {\tool} trained on the PCSD dataset, as well as the time cost {(Note that, since the JCSD dataset has quite similar results to the PCSD dataset, we only exhibit the results on the PCSD dataset).} 
%\song{1-2 sentence to justify why only use PCSD}

We can see that enlarging the maximum deliberation number $K$ generally increases the performance of {\tool}. When enlarging $K$ from 1 to 5, the BLEU-1 score increases by 6.0\%. We also note that {\tool} with $K=2$ substantially outperforms {\tool} with $K=1$ (i.e one-pass model), which indicates that the deliberation process can greatly improve the comment quality by polishing the previously generated comment. Moreover, for $K$ larger than 3, the performances slowly increase but the time cost rises exponentially. {For example, when enlarging $K$ from 3 to 5, the BLEU-1 score increases by 1\% (0.5 points), while the training time increases by 65\% (26 hours).} Thus, we consider $K=3$ to be a trade-off choice between effectiveness and efficiency.

\subsection{Threats to Validity}
There are four main threats to the validity of our approach.

The first threat to validity is that {\tool} uses the lexical similarity based method to retrieve the top similar code-comment pair, which may cause the retrieved comment (initial draft) to be semantically different from the target comment. However, the threats can be largely relieved as {\tool} generates the target comment by iteratively polishing the previous comments. Specifically, {\tool} can correct and refine the retrieved comment in subsequent iterations by leveraging its global information and semantic features of the source code.
Thus, even though the dissimilar comment is retrieved, {\tool} still can guarantee its performance is not affected.

The second threat to validity is the datasets we use. 
We only evaluate {\tool} on the Java and Python datasets. However, {\tool} uses language-agnostic features that can be easily extracted from any programming language. Therefore, we believe that our approach has good generalizability and can perform well on the datasets of other programming languages, such as C\# and Ruby.

The third threat relates to the suitability of evaluation metrics. First, recent researchers have raised concern over the use of BLEU \cite{DBLP:conf/kbse/GrosSDY20}, warning the community that the way BLEU is used and interpreted can greatly affect its reliability. To mitigate that threat, we also adopt other metrics, i.e., ROUGE, METEOR, and CIDEr, when evaluating performance.
{Second, there is also a threat related to our human evaluation. We cannot guarantee that each score assigned to every generated comment is fair. To mitigate this threat, each comment is evaluated by six human evaluators, and we use the average score of the two evaluators as the final score.}

The fourth threat relates to the errors in the implementation of baselines. To mitigate this issue, we directly use the publicly available code of CODE-NN, TL-CodeSum, Hybrid-DRL, Rencos, and Re$^2$com to implement baselines. However, the code of EditSum \cite{DBLP:conf/kbse/LiL000J21} is not available, so we tried our best to understand the paper and re-implement the approach carefully. {While we have verified our implementation can achieve similar results as the original EditSum on the same dataset used in its paper.}

\section{Related Work}
\label{sec:rw}
\subsection{Automatic Comment Generation}
The automatic comment generation task is now a rapidly-growing research topic in the community of software engineering and natural language processing. 

Early studies typically utilize template-based approaches and information retrieval (IR) based approaches to generate comments. The basic idea of the template-based approach \cite{DBLP:conf/kbse/SridharaHMPV10,DBLP:conf/iwpc/MorenoASMPV13,DBLP:journals/tse/McBurneyM16} is to extract the keywords from the code snippets and fill them into the predefined templates. Due to the limitations of manually designing templates, these methods are usually time-consuming and have poor generalization. The IR-based approaches \cite{DBLP:conf/icse/HaiducAM10,DBLP:conf/wcre/HaiducAMM10,DBLP:conf/iwpc/EddyRKC13,DBLP:conf/kbse/WongYT13,DBLP:conf/wcre/WongLT15,DBLP:journals/jasis/DeerwesterDLFH90,DBLP:journals/cacm/SaltonWY75,liu2018neural} aim to use IR techniques to extract keywords from the source code and compose them into term-based comments for a given code snippet. For example, Wong \textit{et al.} \cite{DBLP:conf/wcre/WongLT15} generated a comment for a given code snippet by retrieving the replicated code samples from software repositories with clone detection techniques. However, the IR-based approaches ignore the semantic relationship between source code and natural language, so the comments they generate are poorly readable. Recently, many learning-based methods have been proposed, which train the neural models from a large-scale code-comment corpus to automatically generate comments \cite{DBLP:conf/acl/IyerKCZ16,DBLP:conf/ijcai/HuLXLLJ18,DBLP:conf/kbse/WanZYXY0Y18,DBLP:conf/kbse/WeiLLXJ20,DBLP:conf/iwpc/HuLXLJ18,DBLP:conf/icse/ZhangW00020,DBLP:conf/kbse/LiL000J21,DBLP:journals/tosem/WangXLHWG21,DBLP:conf/icsm/LeClairBM21,DBLP:conf/kbse/ChenZ18}.
Iyer \textit{et al.} \cite{DBLP:conf/acl/IyerKCZ16} first treated the comment generation task as an end-to-end translation problem and introduced NMT techniques into code comment generation. Hu \textit{et al.} \cite{DBLP:conf/iwpc/HuLXLJ18} converted the Java methods into AST sequence to learn the structural information, and applied a seq2seq model to generate comments.
Wei \textit{et al.} \cite{DBLP:conf/kbse/WeiLLXJ20} proposed an exemplar-based comment generation method that utilized the comment of the similar code snippet as an exemplar to assist in generating the target comment.
Zhang \textit{et al.} \cite{DBLP:conf/icse/ZhangW00020} proposed a seq2seq approach that retrieved two similar code snippets for a given code to improve the quality of the generated comment.
Further, Li \textit{et al.} \cite{DBLP:conf/kbse/LiL000J21} treated the comment of the similar code retrieved from a parallel corpus as a prototype. Based on the semantic differences between input code and similar code, they proposed a seq2seq network to update the prototype and generate comments. 

Different from the existing research, we propose a novel framework for automatic comment generation, which performs multiple deliberation processes to iteratively polish the generated comments. {\tool} also contains an evaluation model that not only determines whether to end the deliberation process, but also learns the semantic relationship between source code and target comments. The experimental results also prove the superiority of our approach.

\subsection{Deliberation Networks}
The Deliberation mechanism aims to refine the existing results for further improvement. It has been successfully applied to various domains, such as machine translation \cite{DBLP:journals/corr/abs-1803-05567, DBLP:conf/emnlp/GengF0L18, DBLP:journals/corr/abs-2012-05414}, question generation \cite{DBLP:conf/emnlp/NemaMKSR19}, image captioning \cite{DBLP:journals/corr/abs-2109-08411}, speech recognition \cite{DBLP:conf/icassp/HuSPP20, DBLP:conf/slt/HuPSS21, DBLP:conf/icassp/SungLLL19}.

Xia \textit{et al.} \cite{DBLP:conf/nips/XiaTWLQYL17} first proposed a deliberation network for sequence generation tasks, which consists of two decoders: a first-pass decoder for generating a draft, and a second-pass decoder for polishing the generated draft to a better sequence. Geng \textit{et al.} \cite{DBLP:conf/emnlp/GengF0L18} proposed a novel architecture to introduce the deliberation mechanism into the neural machine translation model. It leveraged the policy network to determine whether to end the translation process adaptively. Nema \textit{et al.} \cite{DBLP:conf/emnlp/NemaMKSR19} utilized the deliberation network to address the automatic question generation task. They proposed a novel approach called Refine Network, which contains two decoders. The second decoder used dual attention to capture information from both (i) the original passage and (ii) the question (initial draft) generated by the first decoder, thereby refining the question generated by the first decoder to make it more correct and complete.
Lian \textit{et al.} \cite{DBLP:journals/corr/abs-2109-08411} proposed a universal two-pass decoding framework for the image captioning task, which contains a drafting model and a deliberation model. The drafting model first generated a draft caption according to an input image, and a deliberation model then refined the draft caption to a better image description. Hu \textit{et al.} \cite{DBLP:conf/icassp/HuSPP20} employed the deliberation network for the speech recognition task. They combined acoustics and first-pass text hypotheses for second-pass decoding based on the deliberation network and obtained significant improvements. 

The findings of previous work motivate the work presented in this paper. Our study is different from the previous work as we focus on enhancing the performance of the comment generation task by incorporating its own characteristics into the deliberation network. Specifically, we combine the two characteristics of the comment generation task into the deliberation network: (1) since code reuse is widespread in software development, we use retrieval techniques to retrieve the most similar comment to provide an explicit hint about the comment expression; (2) since user-defined identifier names usually contain semantic information, we extract the keywords from the source code to strength the semantic features of the source code. 
To the best of our knowledge, this is the first work that treats the comment generation process as the process of writing and polishing, and utilizes multi-pass deliberation automatically generate comments.

\section{Conclusion}
\label{sec:conclusion}
In this paper, we propose a novel multi-pass deliberation framework for automatic comment generation, named {\tool}, which is inspired by human cognitive processes. {\tool} relies on multiple deliberation models and one evaluation model to iteratively perform the deliberation process. For each process, the deliberation model refines the previously generated comment into a better one. The evaluation model estimates the quality of the new generated comment, and compares its quality score to the previous one to determine whether to end the iterative process. We use a two-step training strategy to train our framework.
The evaluation results show that our approach significantly outperforms all other baselines on both Java and Python datasets. A human evaluation study also confirms the comments generated by {\tool} tend to be more readable, informative, and useful. In future work, we plan to incorporate the reinforcement learning techniques (e.g. policy network) into the framework to adaptively choose the suitable deliberation processes, thereby enhancing the performance.
\begin{acks}
We sincerely appreciate anonymous reviewers for their constructive and insightful suggestions for improving this manuscript. 
This work is supported by the National Key Research and Development Program of China under Grant No. 2018YFB1403400, the National Science Foundation of China under Grant No. 61802374, 62002348, 62072442, 614220920020 and Youth Innovation Promotion Association Chinese Academy of Sciences.
\end{acks}

\balance
\bibliographystyle{ACM-Reference-Format}
\bibliography{ref}

\end{document}